\documentclass[aps,twocolumn,floats,nofootinbib]{revtex4} 
\usepackage{graphics,graphicx,epsfig, sidecap} 
\usepackage{amssymb,color} 
\usepackage{epsf,epstopdf,wrapfig} 
\usepackage {amsmath} 
 \usepackage{textgreek} 
\usepackage{sidecap} 
\usepackage{jabbrv} 
\newcommand{\beq}{\begin{equation}} 
\newcommand{\eeq}{\end{equation}} 
\newcommand{\beqn}{\begin{eqnarray}} 
\newcommand{\eeqn}{\end{eqnarray}}

\begin{document} 

\title{Coarse--graining and hints of scaling in a population of 1000+ neurons}

\author{Leenoy Meshulam,$^{1,2,3}$ Jeffrey L. Gauthier,$^{1}$ Carlos D. Brody,$^{1,4,5}$ David W. Tank,$^{1,2,4}$ and William Bialek$^{2,3,6}$} 

\affiliation{$^1$Princeton Neuroscience Institute, $^2$Joseph Henry Laboratories of Physics, $^3$Lewis--Sigler Institute for Integrative Genomics,  $^4$Department of Molecular Biology,  and $^5$Howard Hughes Medical Institute, Princeton University, Princeton, NJ 08544\\
	$^6$Initiative for the Theoretical Sciences, The Graduate Center, City University of New York, 365 Fifth Ave., New York, NY 10016} 

\begin{abstract} 
In many systems we can describe emergent macroscopic behaviors, quantitatively, using models that are much simpler than the underlying microscopic interactions;  we understand the success of this simplification through the renormalization group.  Could similar simplifications succeed in complex biological systems?  We develop explicit coarse--graining procedures that we apply to experimental data on the electrical activity in large populations of neurons in the mouse hippocampus.  Probability distributions of coarse--grained variables seem to approach a fixed non--Gaussian form, and we see evidence of power--law dependencies in both static and dynamic quantities as we vary the coarse--graining scale over two decades.  Taken together, these results suggest that the collective behavior of the network is described by a non--trivial fixed point.
\end{abstract}

\date{\today}

\maketitle

\section{Introduction}

Much of what we know about the brain has been learned by monitoring the electrical activity of individual neurons, one at a time.  But thoughts, actions, percepts, and memories surely result from the coordinated activity of many neurons simultaneously.  Indeed, it has long been hoped the emergence of biological function from interactions among large numbers of neurons might be a genuine collective phenomenon, describable in the language of statistical physics \cite{Cooper1973,Hopfield1982,Hopfield1984,Hopfield1986,Little1974}.  

It can be difficult, however, to connect these theoretical ideas about collective behavior to experiments on single neurons \cite{Amit1992}.
Over the past decade, the experimental exploration of the brain has been revolutionized by the development of methods that make it possible to monitor, simultaneously, the activity of many individual neurons. There has been progress both in direct electrical recording methods \cite{MarreAmodeiDeshmukhEtAl2012,ChungJooFanEtAl2018,JunSteinmetzSiegleEtAl2017,SegevGoodhousePuchallaEtAl2004} and in the development of genetically encodable molecules whose fluorescence responds to electrical activity \cite{Dombeck2010,Harvey2012,NguyenShipleyLinderEtAl2016,GauthierTank2018}.  Although there are important tradeoffs in different experimental methods as applied to different systems, in many cases it now is possible to record from hundreds of cells routinely, and in favorable cases from more than 1000 cells.\footnote{This is a rapidly developing field, and we emphasize that references mentioned are representative rather than exhaustive.}  

How should we think about the dynamics of 1000+ neurons?  Faced with a system that has many degrees of freedom, it is natural to search for a reduced description.  Thus, we describe the flow of fluids not by tracking positions and velocities of all the constituent molecules, but by coarse--grained density, velocity, and pressure fields that reflect averages over large numbers of molecules.  The evident complexity of biological systems has led many people to wonder if similar, systematic coarse--graining could be effective in these cases as well.  

When we pass from molecular dynamics to fluid mechanics we certainly reduce the dimensionality of our description, but this is not a guarantee of simplification.  It is crucial that the equations of motion for the coarse--grained variables are simpler and more universal than the equations describing the microscopic degrees of freedom. In this sense, simplicity in our description of macroscopic behaviors itself is  an emergent phenomenon, and our modern understanding of this emergent simplicity is based on the renormalization group (RG).  

Our goal in this paper is to use the ideas of the renormalization group to analyze experimental data on the patterns of activity in a large network of neurons.   Beyond developing a general strategy, we look in detail at experiments on 1000+ neurons in the mouse hippocampus.   We find that probability distributions of coarse--grained variables seem to approach a fixed form, and we see signs of scaling in both static and dynamic quantities as we vary the coarse--graining scale over two decades.  These results suggest that there is a simpler description of the collective behavior in this network, and that this description is a non--trivial fixed point of the RG.  A preliminary account of these ideas is in Ref  \cite{meshulam+al_18}.

Before proceeding we make two cautionary remarks.  First,  the implementation of the renormalization group is not unique.  There are many different ways of carrying out the coarse--graining step, but the truly universal features of macroscopic behavior are independent of these arbitrary choices.  In the present context,  interesting results emerge from our explicit coarse--graining of neural dynamics, but presumably there is no uniquely best way of doing this coarse--graining.  Second, although we are trying to address a general question about the search for simplification of the neural dynamics, necessarily we look at data from a particular system.  Thus we need to address the connections between the results of our coarse--graining and what we know about the function of the network of neurons in the hippocampus.

\section{RG and data}

To explain our strategy, it is useful to start with a brief summary of the RG and then contrast this theoretical framework with more conventional approaches to data analysis.   The  remarks in this section of course are not a substitute for a full review of either subject, which may be found in classic  \cite{Wilson1971,Wilson1983,Kadanoff1966,Fisher1974} and modern references \cite{Kadanoff2013,Stanley1999,Fisher1998,FahadAlshatriTariEtAlSept.2014,XuTian2015}.  We beg the indulgence of readers who are expert in either subject, and try  to present our ideas in a form that can bridge the gap between the statistical physics community and the ``big data'' community, especially those sharing an interest in the dynamics of biological networks. 

\subsection{Aspects of the RG}
 
The renormalization group implements a widely held intuition, namely that the many interacting degrees of freedom in a system can be divided into one group of interesting variables and another group which we might like to ignore as being ``details.''  In the familiar examples from physics, the details are things that happen on short length scales, while the interesting things are things that we can (sometimes literally)  see on longer length scales.   Crucially, the RG focuses from the beginning on the fact that this division is  arbitrary.  As a result, the important question is not how to define the length scale that marks the ``correct'' boundary between interesting and uninteresting variables, but rather what happens as we move this boundary systematically from one extreme to the other.   

As we average over more and more of the details at short distances, our description of the remaining variables evolves, and it is this evolution---a flow through the space of possible models for the system---that is the output of the RG.  In  successful applications of the renormalization group, the remarkable result is that the flow is toward simpler models.    These simple models, and the structure of the flow itself, predict the observable  collective behaviors,  quantitatively. 

There are two very different ways to implement the ideas of the renormalization group.  One approach is in real space, and the other is in Fourier (momentum) space.  If  the dynamical variables of our model live on a regular lattice, then when we replace each variable by an average with its nearest spatial neighbors we remove one layer of short distance detail.  Alternatively, we can Fourier transform so that variables are labeled not by position $\mathbf x$ but by wavevector or momentum $\mathbf k$; averaging over short distance details then means averaging over the variables with $|{\mathbf k}|$ larger than some cutoff $\Lambda$.  

If we start with variables $\{\sigma_{\rm i}\}$,   coarse--graining in real space means making the replacement
\begin{equation}
\sigma_{\rm i} \rightarrow \tilde\sigma_{\rm i} = f\left( \sum_{{\rm j}\in {\cal N}_{\rm i}}\sigma_{\rm j}\right),
\label{realspace1}
\end{equation}
where ${\cal N}_{\rm i}$ is the spatial neighborhood around variable $\rm i$ over which we average, and  the function $f(\cdot )$ allows for nonlinearities.  As an example, if the original variables are binary, then choosing $f(\cdot )$ to be a step function would insure that the coarse--grained variables remain binary.  Similarly, if we pass to Fourier space, defining
 \begin{equation}
s({\mathbf k}) = {1\over\sqrt{N}}\sum_{\rm i} e^{-i{\mathbf k}\cdot {\mathbf x}_{\rm i}} \sigma_{\rm i} ,
\end{equation}
then we construct coarse--grained variables by limiting the range of wavevectors in the inverse transform,
\begin{equation}
\sigma_{\rm i} \rightarrow \tilde\sigma_{\rm i} =  {{z_\Lambda}\over\sqrt{N}}\sum_{|{\mathbf k}| < \Lambda} e^{i{\mathbf k}\cdot {\mathbf x}_{\rm i}} s ({\mathbf k}) ,
\label{momentumcut}
\end{equation}
where $z_\Lambda$ allows for rescaling.  What is important is that, under these coarse--graining transformations, the joint distribution of variables also is transformed,
\begin{equation}
P\left( \{\sigma_{\rm i}\}\right)  \rightarrow {\tilde P}\left( \{\tilde \sigma_{\rm i}\}  \right) .
\end{equation}
Coarse--graining reduces the number of degrees of freedom, but it is useful to imagine that we can expand the system back to its original size, so that $P$ and $\tilde P$ are distributions over the same number of variables, and hence more directly comparable.   All of the quantitative predictions of the RG emerge from an analysis of the transformation of distributions, $P \rightarrow \tilde P$.  

In equilibrium statistical mechanics, the probability distributions of interest are Boltzmann distributions, 
\begin{equation}
P\left( \{\sigma_{\rm i}\}\right)  = {1\over Z} e^{-\beta H(\{\sigma_{\rm i}\})},
\end{equation}
where $\beta$ is the inverse of the absolute temperature and $H(\{\sigma_{\rm i}\})$ is the Hamiltonian or energy of the system as a function of the microscopic variables.  It then  is natural  to describe the RG as transforming one Hamiltonian into another.   It was emphasized early on by Jona--Lasinio, however, that we can think of the RG more generally as transforming probability distributions \cite{Jona-Lasinio1975}.  This invites us to use RG ideas in a wider range of problems, and gives us a slightly different view of classical results in probability theory such as the central limit theorem.

\subsection{PCA and clustering}

The heart of the renormalization group is a coarse--graining process that reduces the number of degrees of freedom in the system.  The search for such reduced descriptions also is a common feature of modern data analysis methods, in biological systems and beyond.  In many cases the reduction is a matter of convenience, in that we would like to visualize what is happening and we are limited to plotting in two or three dimensions.  More deeply, many data analysis methods aim to find a reduced description which is in some sense optimized, preserving as much as possible of what we find interesting or relevant in the data.  With this optimization, the reduced variables should tell us something about underlying mechanisms.    

Perhaps the dominant approach to dimensionality reduction is principal components analysis (PCA).  The idea is to think of the original variables $\{\sigma_{\rm i}\}$ as living in a vector space, and then find the  Euclidean projection onto a lower dimensional subspace that preserves as much of the total variance as possible.  The solution to this problem is to construct the covariance matrix
\begin{equation}
C_{\rm ij} = \langle \sigma_{\rm i} \sigma_{\rm j}\rangle - \langle \sigma_{\rm i}\rangle\langle \sigma_{\rm j}\rangle ,
\end{equation}
and then find the eigenvalues $\{\lambda_{\rm r}\}$ and eigenvectors $\{ u_{\rm jr} \}$,
\begin{equation}
\sum_{{\rm j}=1}^N C_{\rm ij}u_{\rm jr}  =  \lambda_{\rm r} u_{\rm ir} ,
\label{eigen1}
\end{equation}
with the eigenvalues ordered $\lambda_1 > \lambda_2 > \lambda_3 \cdots $.  As usual we normalize the eigenvectors to unit length, and they are orthogonal, so that
\begin{equation}
\sum_{{\rm r} = 1}^N u_{\rm ir}u_{\rm jr} = \delta_{\rm ij} .
\end{equation}
We can define operators $\hat P (K)$ that project onto the subspace spanned by the $K$ eigenvectors associated with the $K$ largest eigenvalues,
\begin{equation}
{\hat P}_{\rm ij}(K) = \sum_{{\rm r} = 1}^K u_{\rm ir}u_{\rm jr} .
\label{projop_def}
\end{equation}
Finally, coarse--grained variables are defined by these projections,
\begin{equation}
\sigma_{\rm i} \rightarrow \tilde \sigma_{\rm i} = \sum_{{\rm j} = 1}^N {\hat P}_{\rm ij}(K)\sigma_{\rm j},
\label{pcaproject}
\end{equation}
and this is optimal in the sense defined above.

The usual hope in using PCA is that the eigenvalue spectrum will point to a clear reduction of dimensionality.  For this to happen, we need to see that the first $D$ modes capture a large fraction of the total variance, that is $F(D)$ approaches unity for some small $D \ll N$, where
\begin{equation}
F(D) = {
{\sum_{{\rm i}=1}^D \lambda_{\rm i}}
\over
{\sum_{{\rm j}=1}^N \lambda_{\rm j}}
} ,
\end{equation}
and there is a gap in the spectrum between $\lambda_D$ and $\lambda_{D+1}$.  Interestingly, there are several instances in which the {\em output} of the brain passes these  tests for dimensionality reduction, from the crawling movements of the worm {\em C elegans}  \cite{StephensBuenodeMesquitaRyuEtAl2011,StephensJohnson-KernerBialekEtAl2010,StephensJohnson-KernerBialekEtAl2008} to human hand movements \cite{Bizzi1998,SantelloFlandersSoechting1998} and the trajectories of primate eye movements \cite{OsborneHohlBialekEtAl2007,OsborneLisbergerBialek2005}.   It has been very popular to use similar ideas in discussing the patterns of activity in large populations of neurons, but we shall see  that the network that we are studying does not exhibit low dimensional dynamics in this sense.\footnote{The failure of PCA to identify a clear low dimensional structure does not preclude the possibility that the dynamics live on a low dimensional, but curved manifold \cite{low+al_18}.  Only if the embedding dimension of this manifold is small would we see a break in the eigenvalue spectrum.}

An important complement to PCA is the idea of clustering.  While PCA appeals to a geometry in the original space of variables, and explicitly considers rotations in this space as meaningful, clustering takes seriously the discrete identities of the original variables.  Starting from some measure of similarity, such as the correlation matrix
\begin{equation}
c_{\rm ij} = { {C_{\rm ij}} \over{\sqrt{C_{\rm ii}C_{\rm jj}}}} ,
\label{cij_def}
\end{equation}
clustering seeks to group variables together so that similar variables are in the same group. In some applications it is interesting to ask how many clusters provide a natural grouping of the variables, while in other cases we can imagine iterating, searching first for small clusters and then assembling these into larger clusters, hierarchically.  At each stage of the hierarchy we can define new variables which are representative of the variables inside each cluster, perhaps simply their mean.  In this sense the clustering process also implements a kind of dimensionality reduction.

While there are many useful approximate algorithms, finding globally optimal clusters is, in general,  a difficult computational problem, and even simple versions are NP--hard \cite{MahajanNimbhorkarVaradarajan2012}.  In the same way that the crucial output of PCA is the identity of the $D$ most important dimensions, the crucial output of clustering is the identity of the clusters.   Hierarchical clustering provides a visualization of the original variables as being the leaves of a tree, which has clear appeal in the biological context where we are used to thinking about the evolutionary relatedness of organisms or their DNA sequences. When analyzing neural activity, clustering has proved useful in classifying population of neurons with similar coding properties and in identifying different subtypes of neurons in a mixed population \cite{ChungSurmeierLeeEtAl1986,MorcosHarvey2016,CohenHaeslerVongEtAl2012}.

\subsection{Connections}

The renormalization group is an approach to the analysis of models, while PCA and clustering are approaches to the analysis of data.  Nonetheless, there are important connections.  The first step of the renormalization group is a coarse--graining or dimensionality reduction.  In real space, this coarse--graining is an example of clustering, and as we iterate we organize the original variables into hierarchical clusters.  In the cases that we study most often  in statistical physics, where variables live on a spatial lattice and interact with their near neighbors, finding the best clustering is trivial, since variables necessarily are most similar to their spatial neighbors.  What is important in the real space RG, then, is not the identity of the clusters, but the behavior of the joint distribution of the cluster representatives at different stages of the hierarchy. 

In momentum space, the coarse--graining step is essentially the same as in PCA \cite{BraddeBialek2017}.  For a system with translation invariance, the covariance matrix is diagonalized by passing to Fourier space, and so the principal components of the original fluctuating variables {\em are} the Fourier components,  indexed by wavevector or momentum $\mathbf k$.  The associated eigenvalues $\lambda_{\mathbf k}$ are the Fourier transform of the correlation function, $G({\mathbf k})$, where
\begin{equation}
C_{\rm ij} = {1\over N}\sum_{\mathbf k} e^{i{\mathbf k}\cdot({\mathbf x}_{\rm i} - {\mathbf x}_{\rm j} )} G({\mathbf k}) .
\end{equation}
In many systems, $G({\mathbf k})$ is a decreasing function of $|{\mathbf k}|$, so that the principal components which make small contributions to the total variance are exactly the Fourier components with large $|{\mathbf k}|$, corresponding to short distance details.  Quantitatively, coarse--graining by placing a cutoff at $|{\mathbf k}| = \Lambda$, as in Eq (\ref{momentumcut}), then is  the same as projecting onto the first $K$ principal components, as in Eq (\ref{pcaproject}), with $K \sim \Lambda^d$ for systems in $d$ dimensions.  As with the interpretation of real space coarse--graining as clustering, the problem of finding principal components here is trivial, and their meaning is obvious.  What is important is the behavior of the joint distribution of the coarse--grained variables as we move the boundary $\Lambda$.

To summarize, the coarse--graining steps in the renormalization group are essentially the same as common dimensionality reduction methods for data analysis.  Real space renormalization implements iterated or hierarchical clustering, and momentum shell renormalization implements principal components analysis.  But the RG is much more than clustering or PCA.  Rather than searching for the best dimensionality reduction, the RG analyzes what happens to the distribution of the remaining variables as we move the boundary between the details that we average over and the more macroscopic features that we keep.  Indeed, in the RG analysis of models, each step of coarse--graining is followed by an expansion of the system that restores the original number of variables, so that dimensionality reduction itself is never the goal. The renormalization group is a search for simplicity in the space of models, not in the space of system variables.

When we coarse--grain, the variables that remain are combinations of the original variables.  In the simplest cases---including PCA, momentum shell RG, and real space renormalization without the nonlinearity in Eq (\ref{realspace1})---the coarse--grained variables are just linear combinations of the original variables.  As we coarse--grain more and more, the remaining variables then are linear combinations of more and more of the original variables.  If the correlations in the system are sufficiently weak, then the central limit theorem guarantees that the distribution of coarse--grained variables will approach a fixed, Gaussian form.  Perhaps the most important result of the RG is that the distribution of coarse--grained variables can approach a fixed form that is not Gaussian.

In the analysis of models, we have access (at least approximately) to the entire probability distribution.  When we look at data, all we have are samples.  How then can we follow the ``RG flow'' through the space of probability distributions?  Our strategy follows Binder's approach to the analysis of Monte Carlo simulations \cite{Binder1981}.  Rather than trying to follow the joint distribution, he focused on the distribution of the individual coarse--grained variables, and the evolution of this distribution under successive coarse--graining steps.  If the joint distribution of all the variables approaches a Gaussian, then we will see the individual variables also become Gaussian. Conversely, if the full joint distribution approaches a non--Gaussian fixed point, then the distribution for the individual variables should also approach a fixed non--Gaussian form.\footnote{In principle we could do more than this, in effect using the experimental samples as the input to a full Monte Carlo renormalization group \cite{Ma1976,PawleySwendsenWallaceEtAl1984,Swendsen1979}.  But this requires construction of an explicit model for the underlying distribution, which is complicated in the inhomogeneous systems that we consider here.}

The renormalization group teaches us that the approach to a fixed form of the probability distribution is  associated with scaling behaviors for the parameters of this distribution.  As an example, if our coarse--graining procedure replaces each variable by the sum over $K$ variables, and we see the distribution of these coarse--grained variables approaching a Gaussian at large $K$, then the variance (connected second moment) of the distribution should scale as $M_2(K) \propto K$.  On the other hand, if coarse--graining leads to a fixed non--Gaussian distribution then  the variance should be proportional to a different power, $M_2(K) \propto K^{\tilde \alpha}$.  Both the non--Gaussian fixed point and the scaling behavior are signatures of self--similarity in the underlying correlations, which is not generic.  More subtly, we expect that coarse--grained variables fluctuate more slowly than the original microscopic variables, and in many cases we should also see ``dynamic scaling'' so that the correlation time $\tau_c \propto K^{\tilde z}$.    These scaling behaviors, if they exist, represent a considerable simplification in our description of the system.

\subsection{A strategy}

At each moment in time, the activity of each neuron $\rm i$ in the population that we study is described by a variable $\sigma_{\rm i}$.  We want to coarse--grain these variables and follow the flow of their probability distribution.  The hope is to see that this distribution is approaching some non--trivial fixed form as we average over more and more of the ``microscopic'' details.  In addition, the emergence of a  fixed form for the distribution should be associated with scaling behaviors. Our strategy thus begins with  two approaches to coarse--graining, analogous to real space and momentum space as described above.  We then ask what statistical structures in these coarse--grained variables might point to the joint distribution of activity in the network, $P(\{ \sigma_{\rm i}\})$, being controlled by an underlying fixed point of the renormalization group transformation.

\begin{figure*}
\centerline{\includegraphics[width = \linewidth]{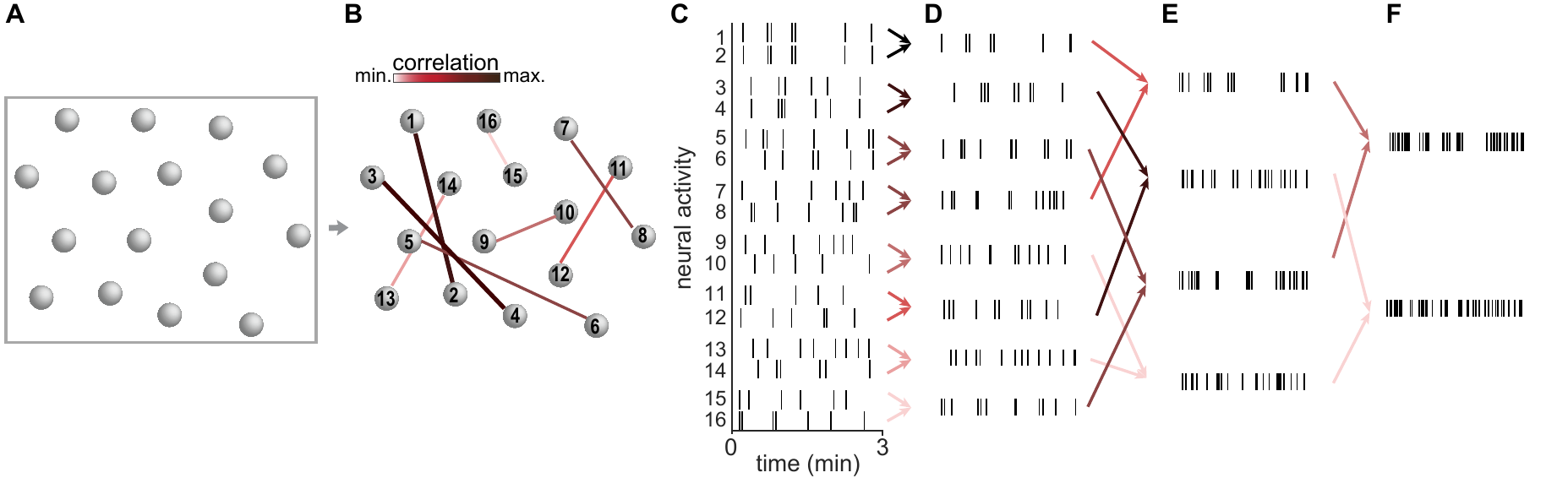}}
\caption{Schematic of our ``real space'' coarse--graining procedure. 
	\textbf{(A)} Field of view of 16 example neurons.
	\textbf{(B)} Initially, computing the hierarchy of correlations strengths between all pairs of neurons based on their activity. Each cell can only participate in one pair. 	
	\textbf{(C)} Activity window for the 16 example neurons, where each row shows the activity a neuron during a time window of three minutes. Every black vertical line indicates a moment where the neuron was active. Then, maximally correlated pairs of neurons are grouped together by summing up the activity in the pairs. Advancing in a greedy fashion in correlation space, all variables will be grouped by the end of each iteration. (Pairs were ordered for ease of visualization).
	\textbf{(D)} The summed up activity of each pair from the previous iteration results in a coarse--grained variable with double the range of activity values of the original neurons. For the next iteration, we find the new correlations between pairs of coarse- grained variables and then group the maximally correlated pairs by summing up their activity.
	\textbf{(E)} Repeat iteration as described in D.
	\textbf{(E)} Repeat iteration as described in D. 
\label{schematic}}
\end{figure*}

{\em Real space, or direct correlations.} In familiar physical systems, we begin coarse--graining by averaging over a small region in space, guided by the fact that interactions are local.   Because neurons are extended objects, capable of connecting across a significant fraction of the region that we are studying, locality is not a strong constraint.  Instead we look at the correlations directly, and coarse--grain by grouping together cells whose responses are most strongly correlated.  We will refer to this as coarse--graining based on direct correlations.  We do this greedily, much as has been done in the analysis of models with strong disorder \cite{Fisher1995}.   Concretely, if we refer to the raw data as $\sigma^{(1)}_{\rm i}$,  we search for the maximal non--diagonal element in the matrix of correlation coefficients $c_{\rm ij}$, then zero the rows and columns associated with this pair of cells ${\rm i},{\rm j}_*({\rm i})$, and iterate.   The result,  shown schematically in Fig \ref{schematic}A and B, is a set of maximally correlated pairs $\{{\rm i}, {\rm j}_*({\rm i})\}$, and we then define coarse--grained variables
\begin{equation}
\sigma^{(2)}_{\rm i} = \sigma^{(1)}_{\rm i} + \sigma^{(1)}_{{\rm j}_*({\rm i})}  ,
\end{equation}
where now ${\rm i} = 1,\, 2,\, \cdots ,\, N/2$.  Importantly, we can iterate this process.  We compute the correlation matrix of the variables $\{\sigma^{(2)}_{\rm i} \}$ and search again for the maximally correlated pairs $\{{\rm i}, {\rm j}_*({\rm i})\}$, then define
\begin{equation}
\sigma^{(3)}_{\rm i} = \sigma^{(2)}_{\rm i} + \sigma^{(2)}_{{\rm j}_*({\rm i})}  ,
\end{equation}
and so on; at each stage we have $N_k = \lfloor N/2^{k-1}\rfloor$ variables remaining.   This coarse graining produces clusters of $K = 2,\, 4,\cdots ,\, 2^{k-1}$ neurons, and the variable $\sigma_{\rm i}^{(k)}$ is the summed activity of cluster $\rm i$. An example of three iterations of this process is shown in Fig \ref{schematic}C-F.

{\em Momentum space, or eigenmodes.}  In systems with translation invariance, we have the choice of coarse--graining either in real space or in Fourier (momentum) space.      We recall that with translation invariance the Fourier modes are the eigenvectors of the covariance matrix, known in other contexts as principal components.   We can put the principal components in order by their contribution to the variance, and then averaging over the components that make small contributions to the variance is analogous to averaging over small variance, short wavelength fluctuations.  More precisely, we coarse--grain by placing a cutoff, keeping only those $\hat K$ principal components that make the largest contributions to the variance, and we refer to this as coarse--graining based on eigenmodes.  Studying what happens to the distribution of the remaining variables as we change the cutoff   then is analogous to the ``momentum shell'' renormalization group \cite{BraddeBialek2017}.   Concretely, we start with the covariance matrix, $C_{\rm ij}$ in Eq (\ref{Cij}), find eigenvalues and eigenvectors as in Eq (\ref{eigen1}), and then define coarse--grained variables as projections  onto the subspace associated with the $\hat K$ largest eigenvalues,
 \begin{equation}
\phi_{\hat K} ({\rm i}) = z_{\rm i}(\hat K) \sum_{\rm j} \hat P_{\rm ij} (\hat K) \left[ \sigma_{\rm i}^{(1)} - \langle \sigma_{\rm i}^{(1)}  \rangle\right],
\label{phiKi}
\end{equation}
the projection operator $P_{\rm ij} (\hat K)$ is defined in Eq (\ref{projop_def}).  In order to track the distributions of these variables as we change our cutoff $\hat K$, it is easiest to subtract the mean, as we have done here, and choose $z_{\rm i}(\hat K)$  so that $\langle\phi_{\hat K}^2 ({\rm i})\rangle  =1$.   

{\em Moments.}     Since our coarse--graining  based on direct correlations involves adding the ``neighboring''  variables, the mean activity increases linearly,
\begin{equation}
M_1(k) \equiv {1\over {N_k}} \sum_{{\rm i}=1}^{N_k} \langle \sigma_{\rm i}^{(k)}\rangle = K M_1(1)  ,
\end{equation}
where after $k$ steps we have $N_k$ clusters each involving $K = 2^{k-1}$ of the original variables.
The first nontrivial question, then, concerns the behavior of the variance in activity,
\begin{equation}
M_2(K) \equiv {1\over {N_k}} \sum_{{\rm i}=1}^{N_k} \left[ \langle \left(\sigma_{\rm i}^{(k)}\right)^2 \rangle - \langle \sigma_{\rm i}^{(k)}\rangle^2 \right]   .
\label{eqvar}
\end{equation}
If our coarse--graining were adding together independent variables, then we would see  $M_2(L) \propto K^1$, while if all variables were perfectly correlated, we would have $M_2(K) \propto K^2$.   If we could see scaling with a nontrivial, intermediate exponent, $M_2(K) \propto K^{\tilde\alpha}$, this would be a hint that correlations  have a self--similar structure and hence a non--trivial fixed point of the RG.

{\em Distributions.} More generally we can look at the full distribution of the individual coarse--grained variables, and ask if this distribution approaches a fixed form as $k$ become large.  When we use the direct correlations,  these variables are the summed activity of neurons inside a cluster of size $K$, so we  separate the probability of complete silence from the distribution of nonzero activity,
\begin{eqnarray}
P(\sigma_{\rm i}^{(k)}) &=& P_{\rm silence}(K) \delta\left(\sigma_{\rm i}^{(k)}, 0\right)\nonumber\\
&&\,\,\,\,\, + \left[1 - P_{\rm silence}(K)\right] F_K (\sigma_{\rm i}^{(k)}/K) ,
\end{eqnarray}
where we define the distribution of the normalized activity $\phi = \sigma_{\rm i}^{(k)}/K$, which measures the fraction of active neurons in the cluster at each moment. As we shall see, $P_{\rm silence}(K)$ has a simple scaling behavior with $K$, and $F_K (\phi)$ approaches a fixed form for large $K$. The analogous question in momentum space is whether the distribution
\begin{equation}
P_{\hat K}(\phi) = {1\over N}\sum_{{\rm i}=1}^N {\bf P}\left[ \phi_{\hat K} ({\rm i}) = \phi \right],
\label{k-spaceP}
\end{equation}
approaches a fixed form as we lower  the cutoff $\hat K$.

{\em Eigenvalues.} In systems with translation invariance, the eigenvalues $\lambda_{\rm r}$ of the covariance matrix correspond to the power spectrum or propagator $G({\mathbf k})$, indexed by the wavevector ${\mathbf k}$.  But at a fixed point of the renormalization group we have 
\begin{equation}
G({\mathbf k}) \propto {1\over {|{\mathbf k}|^{2-\eta}}} .
\end{equation}
Since this is a decreasing function of $|{\mathbf k}|$, the ranking of eigenvalues is from small to large values of this momentum variable, so that for a system in $d$ dimensions we would have rank ${\rm r}\sim |{\mathbf k}|^d$.  Then we would expect to see
\begin{equation}
\lambda_{\rm r} \propto {1\over {{\rm r}^\mu}} ,
\end{equation}
with $\mu = (2-\eta)/d$.  In fact we can say more than this if we examine systems of different sizes, since the largest wavevector is set by the microscopic structure of the system (the lattice spacing) while the smallest wavevector is related to the size of the system.  Hence the largest eigenvalue (lowest rank) should depend on the size of the system, while the smallest eigenvalue (highest rank) should not; the result is that in systems with $K$ degrees of freedom we should see
\begin{equation}
\lambda_{\rm r} = A \left({K\over {{\rm r}}}\right)^{\mu} .
\label{lambda_scale}
\end{equation}
Although it is not a rigorous connection, the analogy between principal components and momentum shells suggests that this scaling behavior of the eigenvalues of the covariance matrix should be a signature of a fixed point in the RG flow of probability distributions, but there are two cautionary notes.  First, it is crucial that when we consider systems of different sizes we not take random fractions of a larger system, but rather do something analogous to looking at spatially contiguous regions; fortunately, such clusters are exactly what we construct through our ``real space'' coarse graining.  Second, in practice it might not be possible to observe scaling as in Eq (\ref{lambda_scale}) over a very wide dynamic range.\footnote{The variance of coarse--grained variables, Eq (\ref{eqvar}), is the sum over all elements in the correlation matrix of the raw variables that belong to a cluster.  This quantity is not an invariant, and need have no simple connecting to the spectrum of eigenvalues.  In systems where the raw variables live on a lattice and correlations are translation invariant, one can make such a connection, and hence the exponents $\tilde \alpha$ and $\tilde\mu$ would be related, but we don't expect this to be true in general.}  

{\em Dynamic scaling.}  Establishing correlations requires information to be propagated through the system, and this takes time.  If correlations are self--similar, it is reasonable to expect some corresponding  self--similarity in the dynamics.  With local interactions, this idea is quantified in the dynamic scaling hypothesis---dynamics on a length scale $L$ should be associated with a time scale $\tau \propto L^z$---which we know to be correct for a wide range of models near their critical points.   As emphasized by Cavagna et al \cite{CavagnaContiCreatoEtAl2017}, dynamic scaling provides an independent path to testing for critical behavior in complex biological systems, where many of the usual statistical physics tools are unavailable.   Concretely, if we compute the normalized correlation function of the coarse--grained variables,
\begin{equation}
C^{(k)}_{\rm i}(t) = {{\langle \sigma^{(k)}_{\rm i} (t_0) \sigma^{(k)}_{\rm i} (t_0 +t)  \rangle - \langle \sigma^{(k)}_{\rm i}\rangle^2}\over{\langle [\sigma^{(k)}_{\rm i}]^2 \rangle - \langle \sigma^{(k)}_{\rm i}\rangle^2}} ,
\label{modecorr}
\end{equation}
dynamic scaling means that the behaviors at different levels of coarse--graining will collapse onto a single curve,
\begin{equation}
C^{(k)}(t) = C[t/\tau_c(k)] ,
\end{equation}
with $\tau_c (k) \propto K^{\tilde z}$.  Similarly, if we project onto the individual principal components or eigenmodes,
\begin{equation}
\tilde\sigma_{\rm r}(t) = \sum_{{\rm i}=1}^N \sigma_{\rm i} u_{\rm ir},
\end{equation}
with  $\langle (\delta\tilde\sigma_{\rm r})^2\rangle = \lambda_{\rm r}$, then the normalized correlation functions 
\begin{equation}
C^{\rm r}_{\rm modes}(t) = {1\over{\lambda_{\rm r}}}
\left[ \langle \tilde\sigma_{\rm r} (t_0) \tilde\sigma_{\rm r}(t_0 +t)  \rangle - \langle \tilde\sigma_{\rm r}\rangle^2  \right],
\end{equation}
should obey
\begin{equation}
C^{\rm r}_{\rm modes}(t) = \tilde C[t/\tau_c({\rm r} )] .
\end{equation}
Finally, we should see $\tau_c({\rm r} ) \propto \lambda_{\rm r}^{{\tilde z}/\mu}$

{\em Remarks.}  In a full theory, the exponents that we have defined ($\mu ,\, \tilde\alpha ,\, \tilde z$) should be related to one another and to the form of the distribution that emerges through coarse--graining, all calculable from some effective theory.   But asking for a theory  is getting ahead of ourselves.  We are trying to use renormalization group ideas very far from the domain in which they have been well tested, and we have no guarantee that this approach will work.  The possibility that activity in real networks of neurons can be described by a non--Gaussian fixed point, associated with nontrivial scaling behaviors, is far from obvious.  Our goals here are phenomenological, exploring what happens as we coarse--grain our description of activity in a network of real neurons, and searching for hints of scaling that could point the way toward a deeper theory.   

\begin{figure}
\centerline{\includegraphics[width=\linewidth]{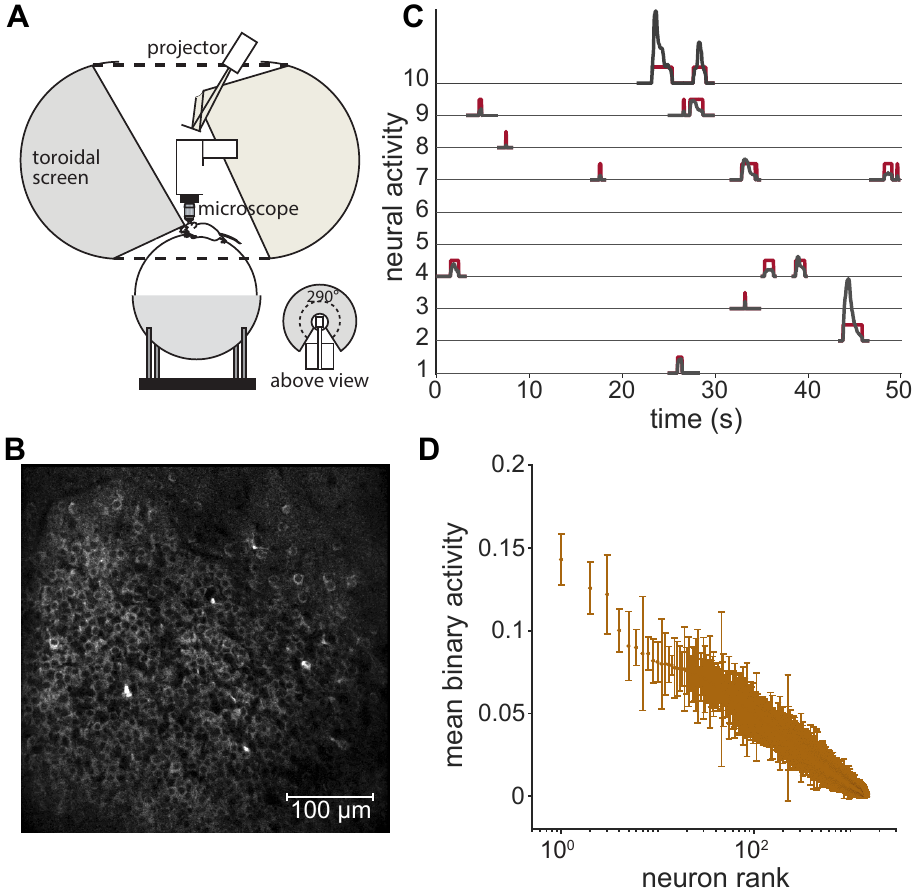}}
\caption{\textbf{Experimental setup and data.} \textbf{(A)} Experimental setup consists of a styrofoam wheel whose motion advances the position on a virtual corridor projected on a surrounding toroidal screen, a head fixed mouse running on the ball, and a two photon microscope which enables {\textit {in vivo}} cellular resolution imaging, as in Ref \cite{GauthierTank2018}. \textbf{(B)} Fluorescence image of neurons in the hippocampus expressing the calcium sensitive protein GCaMP3. \textbf{(C)} Fluorescence signals from 10 randomly selected neurons from the image shown panel B. In gray are identified activity events in the continuous signals after denoising, and in red we show the on/off discretization into binary variables, following Ref \cite{MeshulamGauthierBrodyEtAl2017}. \textbf{(D)} Mean activity of each individual neuron, $\langle \sigma_{\rm i}\rangle$, shown in rank order.  Error bars are the standard deviations across random quarters of the data. 
\label{data}}
\end{figure}

\section{A population of real neurons}

The experiments that we analyze here involve monitoring the activity of 1000+ neurons in a hippocampus of a mouse.  The hippocampus, especially in rodents, is one of the most thoroughly studied regions of the mammalian brain.   Most famously, almost fifty years ago O'Keefe and Dostrovsky discovered that there are individual neurons in the hippocampus that are active only when the animal visits a particular place in the environment \cite{OKeefe1971}.  These are called ``place cells,'' and the activity in a large population of place cells provides a ``cognitive map'' that plays a crucial role in the animal's ability to navigate \cite{PearceRobertsGood1998,OKeefe1998,OKeefe1971}.  There are controversies about the precise role of the hippocampus in humans and other primates, where it has been implicated in episodic memory, among other functions \cite{HunsakerLeeKesner2008,BurgessMaguireOKeefe2002,TulvingMarkowitsch1998,Rolls1996,Eichenbaum2000}.  Recent work shows that, even in rodents, hippocampal neurons can be selective for compact regions in more abstract spaces \cite{EichenbaumKupersteinFaganEtAl1987,AronovNeversTank2017,Sakurai2002}.  

We make use of methods that have been developed over several years \cite{HarveyCollmanDombeckEtAl2009,Dombeck2010,GauthierTank2018}.  As schematized in Fig \ref{data}A, a mouse runs on a styrofoam ball that is levitated on a column of air; the mouse's head is fixed, and the motions of the ball are used to drive a virtual reality system, in this experiment simulating a run along a virtual linear track.  The mice have been genetically engineered to express GCaMP3, a calcium--sensitive fluorescent protein, and this fluorescence is measured with a scanning two--photon microscope as the mouse runs. The microscope is focused on the CA1 region of the hippocampus, where cells form a nearly a single densely packed layer. Figure \ref{data}B shows one image of the neurons in our field of view, and Fig \ref{data}C shows examples of the fluorescence signal vs time from several cells in this population of $N=1485$ cells; pixels are $0.87\times 0.87\,\mu{\rm m}$, and the field of view  is $433 \times 433 \,\mu{\rm m}^2$.    Details of the experiment and the preprocessing of the data are described in Ref \cite{GauthierTank2018}.

The fluorescence signal is sparse, and is denoised to recover a flat baseline, as shown in Fig \ref{data}C.  Nonzero transient signals are brief, but still longer and smoother than the underlying bursts of action potentials, reflecting both the way in which calcium concentration follows electrical activity and the dynamics of GCaMP3 itself; we record in frames of $\Delta t = 1/30\,{\rm s}$, continuously for $40\,{\rm min}$.  We note that these reporter molecules are undergoing constant development, and it is reasonable to expect that imaging experiments of the sort described here soon will be able to resolve individual action potentials.  For now, however, we take the fluorescence signals at face value as correlates of electrical activity.  In the simplest case, we discretize  so that cells are either ``on'' or ``off'' in each frame, $\sigma_{\rm i} \equiv \{0,1\}$, following our previous work \cite{MeshulamGauthierBrodyEtAl2017}; we return to the raw, continuous signals below \cite{meshulam+al_18}.

We start by summarizing basic features of the data.   In Fig \ref{data}D we show the mean of each binary activity variable, $\langle \sigma_{\rm i}\rangle$.  As expected from Fig \ref{data}C, the activity is sparse, so that almost all neurons are active less than $10\%$ of the time, and the distribution extends to very low activity. This long tail of weak activity is in some ways expected.  In this dataset, $\sim 50\%$ of the neurons are place cells, with one or more spatial fields in the environment. But any single experiment samples only a limited environment, and it is likely most neurons will not be place cells in that particular environment; there is controversy about the role that these ``non--place'' neurons could play.  Our recent work with populations of $\sim 100$ neurons shows that place cells and non--place cells can be described on equal footing as part of the collective activity in the network \cite{MeshulamGauthierBrodyEtAl2017}.  More generally, traditional methods of recording from one neuron at a time are naturally biased toward finding more active neurons, so  that improved techniques for recording from more neurons simultaneously reveal the tail of low mean activity in any network.

\begin{figure}
	\includegraphics[width=\linewidth]{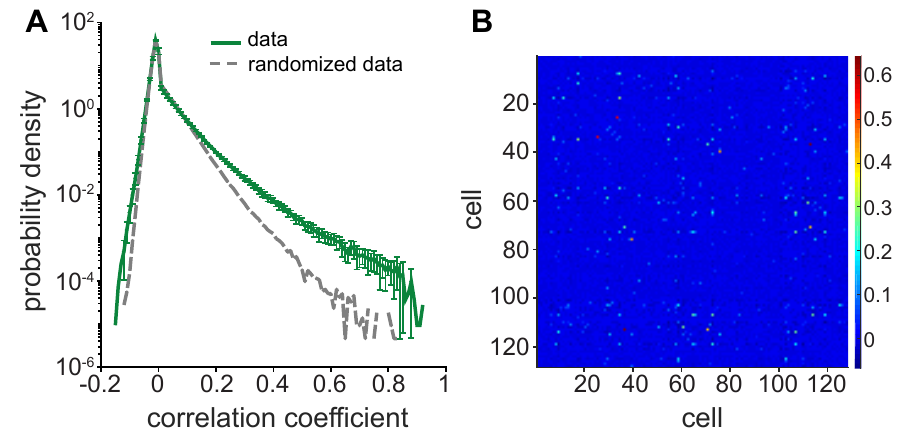} 
	\caption{\textbf{Pairwise correlations.} \textbf{(A)} Distribution of pairwise correlation coefficients.  Mean and standard deviations across random quarters of the experiment (orange), compared with results from randomized data (dashed gray).
		\textbf{(B)} Matrix of correlation coefficients for 128 randomly chosen neurons.  The diagonal has been set to zero, as have matrix elements which are not separated from zero by more than one error bar. 
	}	
	\label{corrs}
\end{figure}

Our coarse--graining procedures depend crucially on the correlations between cells.  In Fig \ref{corrs} we show the  correlation coefficients $c_{\rm ij}$ for the binary variables $\sigma_{\rm i}$; since we have more than one million distinct pairs it makes sense to think of these correlations as coming from a distribution.  For comparison, we have randomized the data, shifting the time axis for each neuron independently; this should break correlations between neurons, but preserves the temporal correlations for each cell.  The mean correlation coefficient for the randomized data is very nearly zero ($\sim 10^{-5}$), as it should be, but the standard deviation is $\delta C = 0.03$, which is consistent with there being at most a few thousand independent samples.\footnote{We can compare the correlation matrix of the randomized data with truly random matrices, and in this way make a more precise estimate of the number of independent samples.  Details will be given elsewhere.}

The degree of overlap between the correlation coefficients in the real and randomized data means that we need to proceed carefully.  We emphasize that this is likely to be a feature of most experiments on large populations of neurons, unless it becomes possible to lengthen recording times in proportion to the increasing numbers of cells being monitored.  Quantitatively, 48\% of the correlation values in the data are large enough in absolute value that they are more likely to occur in the real data than in randomized data.  Similarly, if we attach errors to the measurement of each element in the correlation matrix, then half ($0.55$) have error bars that do not overlap zero.   Importantly, our coarse--graining procedure depends only on the tail of highly significant correlations.

An additional reason to exercise caution is the sensitivity of correlation estimates to misattribution of the non--zero events (Fig \ref{data}B)  to the wrong cell during the initial pre--processing stages of the raw image data. Activity from an overlapping cell can contaminate the original time series, and result in a biased estimate of correlation between the cells \cite{gauthier+al_18}. To reduce our susceptibility to this problem we have examined most transients manually for correct assignment.  All fluorescent shapes underlying transients that resembled the cell were kept, but transients whose underlying shape resembled a different cell or a mix of the neuron and a neighboring one (rare), were discarded.

To illustrate the correlations more explicitly, we show in Fig \ref{corrs}B a segment of the full correlation matrix for 128 randomly chosen neurons.  Although details vary across different systems, the pattern here is familiar---correlations have both signs, and are weak but widespread. More quantitatively, the range of correlations we see in the hippocampus is similar to that reported for mouse primary visual cortex and primate primary motor cortex, and significantly stronger than observed in the salamander retina \cite{DenmanContreras2014, MaynardHatsopoulosOjakangasEtAl1999, SchneidmanBerryIISegevEtAl2006}.  We expect that cells with overlapping place fields will be active simultaneously, and hence positively correlated, while cells with non--overlapping place fields will never be active together, and hence negatively correlated; this is qualitatively consistent with the data, but we have emphasized that place fields alone do not provide a quantitative account of the global properties of the correlation matrix \cite{MeshulamGauthierBrodyEtAl2017}.  The fact that correlations extend through the entire network is reminiscent of  mean--field models, but note that in mean--field we expect correlations on the order of $1/N$ or $1/\sqrt{N}$; with $N>1000$ neurons we have $1/\sqrt{N}\sim 0.03$, which could not be reliably distinguished from zero in this data set.  We conclude that there are widespread correlations, but with magnitudes substantially larger than expected in mean--field theory.

\begin{figure}
	\includegraphics[width=\linewidth]{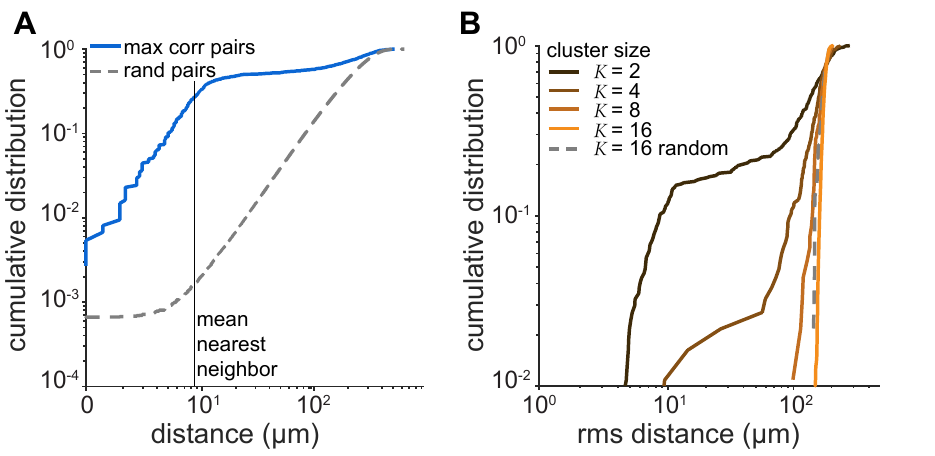}
	\caption{\textbf{Pairwise distances.}  
		\textbf{(A)} Distances between maximally correlated pairs that are summed in the first stage of coarse--graining.  Shown as a cumulative distribution (green, fraction of pairs with distance less than value on the x--axis), and compared with random pairings (dashed gray).  
		\textbf{(B)} Distances among cells in successive stages of coarse--graining.  We measure the root-mean-square distance among cells in the clusters of different sizes $K$, and plot the results as a cumulative distribution.  At $K=16$ we compare with randomly chosen cells. 
	}	
	\label{distances}
\end{figure}

Although we have emphasized that the search for maximally correlated pairs does not rely on spatial neighborhoods, it nonetheless is interesting to ask about the physical locations of the cells that are grouped together in the process of coarse--graining.  The experiment involves observing cells in a $\sim 0.5\times0.5\,{\rm mm}^2$ region; the mean distance to a nearest neighbor is $d = 7.75\,\mu{\rm m}$, indicating that some of the cells are partially overlapping, while individual cells have dendritic arbors that allow them to make contact with cells over a radius of more than one hundred microns\footnote{Note that all the images are in a single plane, and the projection into this plane can lead to overlaps.  As noted above we take considerable care to disentangle the signals in such cases.  The location of a cell is defined as the center of mass (fluorescence intensity) of the sometimes irregular  region assigned to that cell.} Nonetheless, as we see in Fig \ref{distances}A, a significant fraction of cells are paired with close neighbors in the first step of coarse--graining;  $21\%$ of the cells are paired with their nearest neighbors, and roughly half of the cells are paired with cells within $2\times d$. While the first step of coarse--graining often is close to averaging in space, this is not true for subsequent iterations.  We see in Fig \ref{distances}B that the distribution of root-mean-square distances among cells in a cluster changes significantly as we look at $K=4$, $8$, and $16$ cell clusters; at $K=16$ the distribution is almost the same as  when we choose cells at random.  Thus, as expected, spatial locality is not a good guide to coarse--graining. 

\section{Results of coarse--graining via direct correlations}

In this and the following Section, we walk through the analysis of a single data set, forty minutes in the life of one mouse running along a virtual track.  We then ask about the reproducibility of our results in \S \ref{repro}.

\begin{figure*}
	\includegraphics[width=\linewidth]{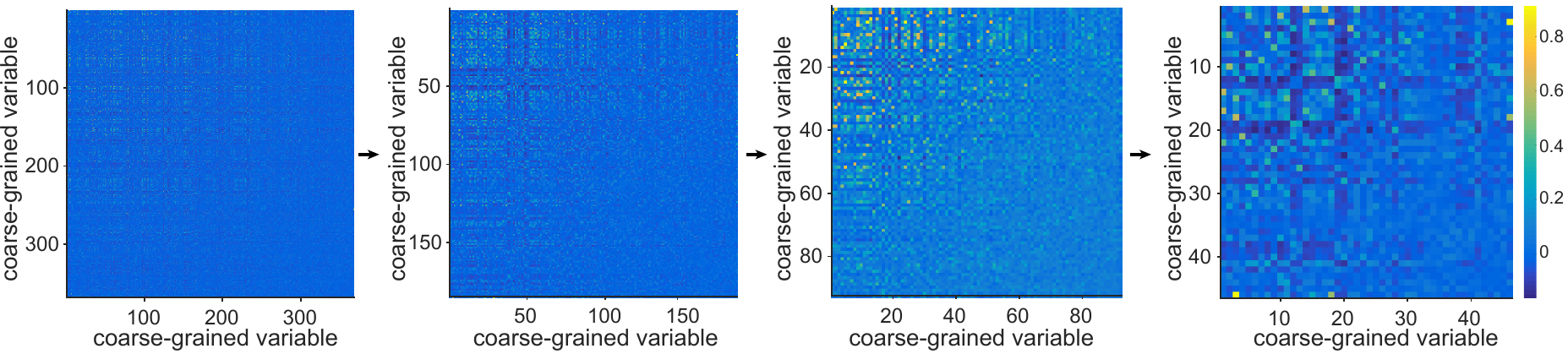} 
	\caption{\textbf{Evolution of correlations through the coarse-- process.} 
		Evolution of correlations between pairs of coarse grained variables, for cluster sizes$ K= 4,\, 8 ,\, 16,\, 32$ (left to right). The diagonal has been set to zero.	Neurons were ordered such that the ``place--cells'' are ordered first and then the rest of the cells. Correlations structure between the place fields becomes gradually more pronounced during the coarse--graining process.\vspace{20 pt}
	\label{evolution_corrs}	
	}	
\end{figure*}

Our coarse--graining procedure groups together pairs of variables based on the strength of correlation between them, in an iterative fashion.  Figure \ref{evolution_corrs} displays the evolution of correlation between neurons through the iterations of coarse-graining. The first panel displays the correlations after the neurons were already grouped twice, for cluster size $K = 4$ , yet it is still difficult to see the ``neighborhoods'' that are driving the coarse--graining. As we increase to $K = 8$ and $K = 16$, the pattern gradually becomes more pronounced. If this were a system with a finite correlation length, we would have expected the blocks formed through the coarse--graining process to eventually become independent of each other; instead, we see the coarse--grained variables remain correlated with each other, and the global structure becomes sharper.

\subsection{Scaling in the variance of cluster activity}

\begin{figure}[b]
	\centering
	\includegraphics[width = \linewidth]{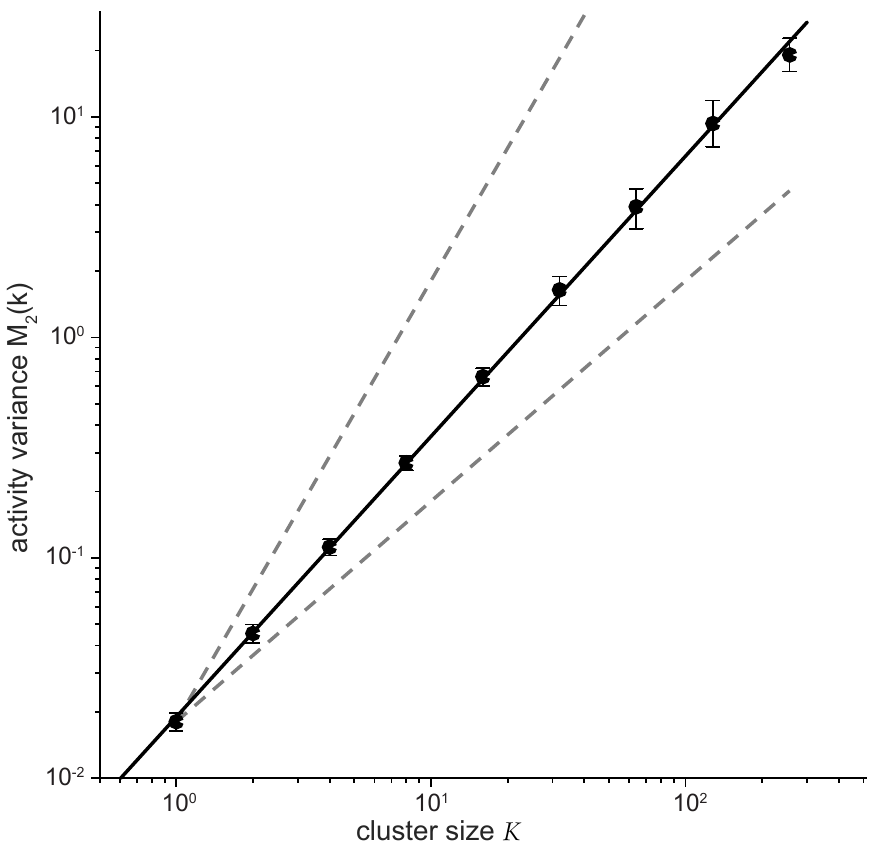}
	\caption{\textbf{Scaling in the variance of activity.} Variance of coarse--grained activity vs cluster size, from Eq (\ref{eqvar}).  Error bars estimated from random connected quarters of the data, shown at four times actual size. Black line indicates a fit,  $M_2(K) \propto K^{\tilde\alpha}$, with $\tilde\alpha = 1.4\pm0.06$.  Dashed lines are $\tilde\alpha = 1$ and $\tilde\alpha =2$, for comparison. 
	\label{var}
	} 
\end{figure}

Our coarse--graining procedure groups neurons into clusters, and defines variables that are the summed activity within each cluster.  How do the statistics of these variables depend on cluster size $K$?  We recall that if we group together independent neurons then the variance $M_2$  in summed activity [Eq (\ref{eqvar})] grows linearly with $K$, while if neurons in a cluster are perfectly correlated then the variance grows quadratically.  In Fig \ref{var} we show $M_2(K)$. What we see is almost perfect scaling, $M_2(K) \propto K^{\tilde\alpha}$, with $\tilde\alpha = 1.4\pm0.06$, across two decades in cluster size. The observation of scaling with a nontrivial, intermediate exponent is a hint that correlations in this network have a self--similar structure.  

In what follows it will be important to have error bars on our estimates of exponents.  This problem is connected with the larger issue of identifying power--law behaviors, a topic that has generated considerable controversy \cite{ClausetShaliziNewman2009}, and to the reproducibility of scaling behavior across experiments, discussed in \S \ref{repro}.   Error bars for all scaling behaviors reported in this work were estimated as the standard deviation across quarters of the data. To respect temporal correlations, the location of the quarter was chosen at random, but time points remained in order inside it, i.e. a connected quarter.  For each quarter of the data, exponents are estimated as the slope of the best fit line on a log--log scale.  The fact that our estimated error in $\tilde\alpha$ is a bit less than 5\% is connected to the fact that the data really do fit very well to a power--law, and the errors on individual points are quite small.\footnote{Caution is required in evaluating the quality of the fit.  In particular, since groups of size $K$ are built out of smaller groups, errors on properties measured at different $K$ are correlated.  This is why we use an empirical measure of error in the estimates of exponents, rather than propagating errors on individual data points.}

\subsection{Silence and activity}

If the  power--law growth of variance is a sign of genuine self--similarity, then we should be able to see related signatures in other aspects of the data.  Suppose that we try to write the full joint probability distribution for the variables inside a cluster of size $K$:
\begin{eqnarray}
P_K \left(\{\sigma_{\rm i}^{(1)}\}\right) &=& {1\over{Z_K}} \exp\left[- E \left(\{\sigma_{\rm i}^{(1)}\}\right)\right],
\label{boltz1}\\
E \left(\{\sigma_{\rm i}^{(1)}\}\right) &=& \sum_{{\rm i}=1}^K h_{\rm i} \sigma_{\rm i}^{(1)} 
+  \sum_{{\rm i},{\rm j}=1}^K J_{\rm ij} \sigma_{\rm i}^{(1)}\sigma_{\rm j}^{(1)} 
\nonumber\\
&&
+  \sum_{{\rm i},{\rm j},{\rm k}=1}^K G_{\rm ijk}  \sigma_{\rm i}^{(1)}\sigma_{\rm j}^{(1)} \sigma_{\rm k}^{(1)} +\cdots ,
\label{boltz2}
\end{eqnarray}
which is a Taylor series in the dependencies or interactions among the neurons.  Since the basic activity variables $\sigma^{(1)} = \{0,1\}$, the probability of complete silence in the cluster is given by
\begin{equation}
P_{\rm silence} \equiv P_K \left(\{\sigma_{\rm i}^{(1)} = 0\}\right)  = {1\over{Z_K}} .
\end{equation}
But Eq (\ref{boltz1}) is a Boltzmann distribution, with units of ``energy'' such that $k_BT = 1$, and so it is natural to identify the effective free energy $F(K) = -\ln Z_K = \ln P_{\rm silence}$.   Complete silence corresponds to the coarse--grained variable $\sigma^{(k)}=0$, which allows us to estimate $P_{\rm silence}$, and hence the free energy, directly \cite{TkacikMarreAmodeiEtAl2014}.

\begin{figure}[t]
	\includegraphics[width=\linewidth]{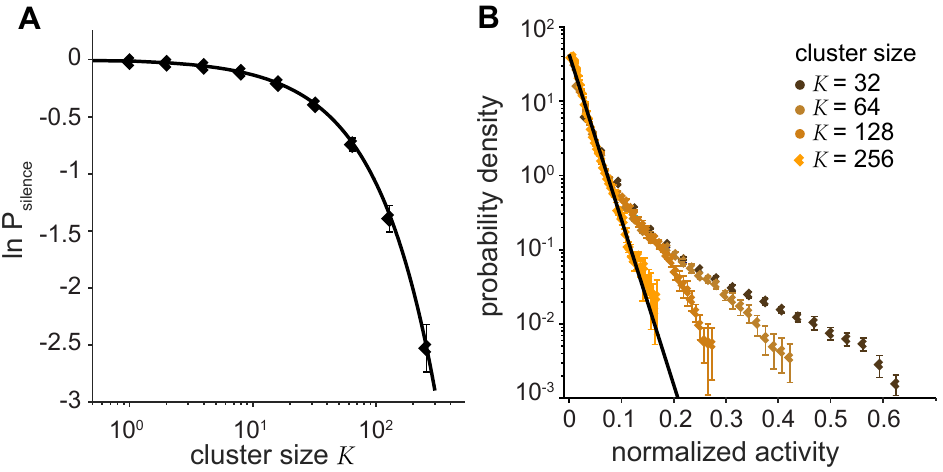}
	\caption{{\textbf{Scaling in the probability of silence and activity.}} \textbf{(A)} Probability of silence in clusters of different sizes.  Error bars are standard deviation across random quarters of the data, and the curve is from the effective free energy $F = -a K^{\tilde\beta}$, with $\tilde\beta = 0.88\pm 0.01$ [Eq (\ref{FvsK})].
	\textbf{(B)} Distribution of nonzero activity in clusters of size $K=32$ (dark brown), $64$ (mid brown), $128$ (light brown), and $256$ (red); line is the exponential distribution. Error bars are standard deviations across random quarters of the data.
	\label{PsilencePact}
}
\end{figure}	

In Figure \ref{PsilencePact}A we show the effective free energy vs cluster size.  The data fit very well to a scaling form 
\begin{equation}
F = -a K^{\tilde\beta} ,
\label{FvsK}
\end{equation}
with the exponent $\tilde\beta = 0.88 \pm 0.01$. Naively one might expect that the probability of silence in a population of $K$ neurons declines exponentially with $K$, which corresponds to extensive growth of the free energy.  Here we see slightly subextensive growth.  If this scaling behavior persists, it would imply  that the free energy per neuron vanishes as $K\rightarrow \infty$.\footnote{There is an interesting connection to results on activity in populations of ganglion cells from  the vertebrate retina \cite{tkacik+al_15}. Both in the raw data and in maximum entropy models that describe these data, it was possible to estimate the microcanonical entropy as a function of the ``energy,'' or log probability.  As the populations became larger, the entropy vs energy approached a line of unity slope, $S(E) \rightarrow E$.  If this pattern continues as $N\rightarrow\infty$, then the free energy per neuron vanishes, as seen here for hippocampal neurons.  }

As noted above, the coarse--grained variable $\sigma_{\rm i}^{(k)}$ corresponds to the summed activity in a population of $K= 2^{k-1}$ cells.  We can think of this as measuring the fraction $x= \sigma_{\rm i}^{(k)}/K$ of cells which are active, and in  Fig \ref{PsilencePact}B we show  the distribution of nonzero $x$ for $K=32,\, 64,\, 128,\, 256$.  We see that the bulk of the distribution is almost invariant under coarse--graining, and that the tail is converging toward an exponential distribution.

\begin{figure}[b]
	\centerline{\includegraphics[width = 0.9\linewidth]{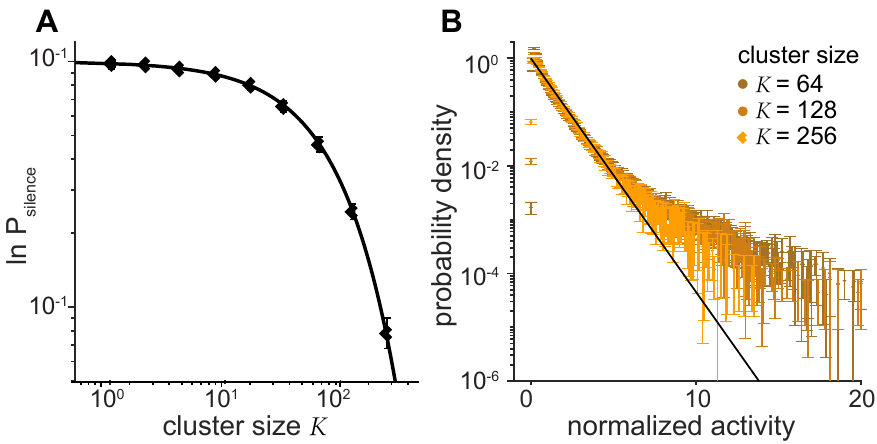}}
	\caption{{\bf{Coarse--graining of continuous activity variables -- scaling in the probability of silence and activity.}} 
		\textbf{(A)} Probability of observing an inactive cluster, $x^{(k)}\equiv 0$, as function of the cluster size $K$.  Results are as in Fig \ref{PsilencePact}A, with $\tilde\beta = 0.893\pm 0.003$.  
		\textbf{(B)} Distribution of nonzero activity, in clusters of size 64 (dark brown), 128 (light brown), and 256 (orange);   error bars computed across random fractions of the data as at left.  The line is $P(x) = e^{-x}$.
		\label{scaling_cont}}
\end{figure}

We can perform the same analysis on the original, continuous version of the data, before binarization \cite{meshulam+al_18}.   The difficulty is that absolute signal amplitudes may vary systematically from cell to cell, for example because of variations in the expression level of the indicator molecule. To compensate for such variations, we normalize the continuous signal $x_{\rm i}$ so that the nonzero values in each neuron have mean one, and we maintain this at each stage of coarse--graining.   We implement coarse--graining of the continuous activity signals as before, searching greedily for maximally correlated pairs $\{{\rm i}, {\rm j}_*({\rm i})\}$ and defining
\begin{equation}
	x^{(k+1)}_{\rm i} = z_{\rm i}^{(k)}\left[x^{(k)}_{\rm i} + x^{(k)}_{{\rm j}_*({\rm i})}  \right],
\end{equation}
where $z_{\rm i}^{(k)}$ enforces the (re)normalization described above.  Note that the maximally correlated pairs determined from the continuous variables may be different than those determined from the discrete variables.  Thus, even though $x\equiv 0$ and $\sigma\equiv 0$ are the same condition in the raw data, these could in principle diverge as we coarse--grain.  In Fig \ref{scaling_cont}A we see that,  $P(x=0) \propto \exp(-a K^{{\tilde\beta}})$, with the same $\tilde\beta$ as in Fig \ref{PsilencePact}A, suggesting that our discretization preserves, quantitatively, the structure of correlations in the system.  Similarly, if we examine the distribution of nonzero values for $x$ (Fig \ref{scaling_cont}B), we see essentially the same behavior as in Fig \ref{PsilencePact}B, perhaps even more clearly. 

\subsection{Eigenvalue spectra}

To explore further, we would like to look directly at the correlations, rather than at their integrated consequences for the coarse--grained activity.  The difficulty, as noted above, is that we have only a few thousand independent samples, and so it is dangerous to ask directly about global properties of the correlations for all $N = 1485$ variables in our data set.  But our coarse--graining procedure identifies clusters of $K = 2,\, 4,\, \cdots ,\, 32,\, 64,\, 128$ cells, analogous to contiguous regions  in systems with spatially local interactions.  Within each cluster, up to $K=128$,  the data are in the regime where we have $>10\times$ more samples than dimensions.  

\begin{figure}
	\centering
	\includegraphics[width = 0.9\linewidth]{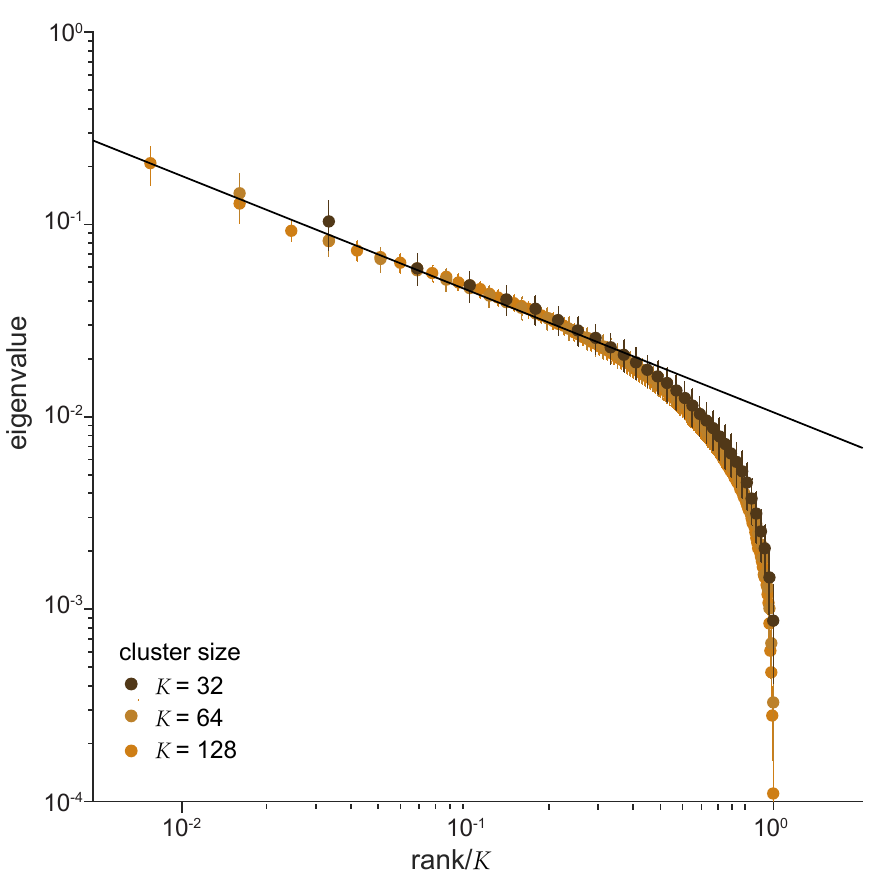}
	\caption{{\textbf{Scaling in the eigenvalue spectra.}} Eigenvalues of the covariance matrix [Eq (\ref{Cij})] for clusters of size $K=32$ (dark brown), $64$ (mid brown), and $128$ (light brown).  Error bars are standard deviations across both random fractions of the experiment and the different clusters in our data set.  Solid line is Eq (\ref{lambda_scale}), with $\mu = 0.71\pm 0.06$. 
		\label{spectra} 
	}

\end{figure}

For all the cells inside one cluster, we can estimate the covariance matrix
\begin{equation}
C_{\rm ij} \equiv \langle \sigma_{\rm i}^{(1)} \sigma_{\rm j}^{(1)}\rangle 
- \langle \sigma_{\rm i}^{(1)}\rangle\langle   \sigma_{\rm j}^{(1)}\rangle  ,
\label{Cij}
\end{equation}
and we can compute the eigenvalues $\lambda_{\rm r}$ of this matrix, which we arrange in decreasing order, as shown in Fig \ref{spectra}.    We see a power--law behavior, as in Eq (\ref{lambda_scale}), admittedly over a limited range, with  $\mu = 0.71\pm 0.06$. Notice that these spectra exhibit scaling in two senses:  the eigenvalues have a power--law dependence on rank ${\rm r}$, and the spectra in clusters of different size $K$ depend only on relative  rank ${\rm r}/K$.
 
\subsection{Dynamic scaling}

\begin{figure*}
	\centering
	\includegraphics[width=0.9\linewidth]{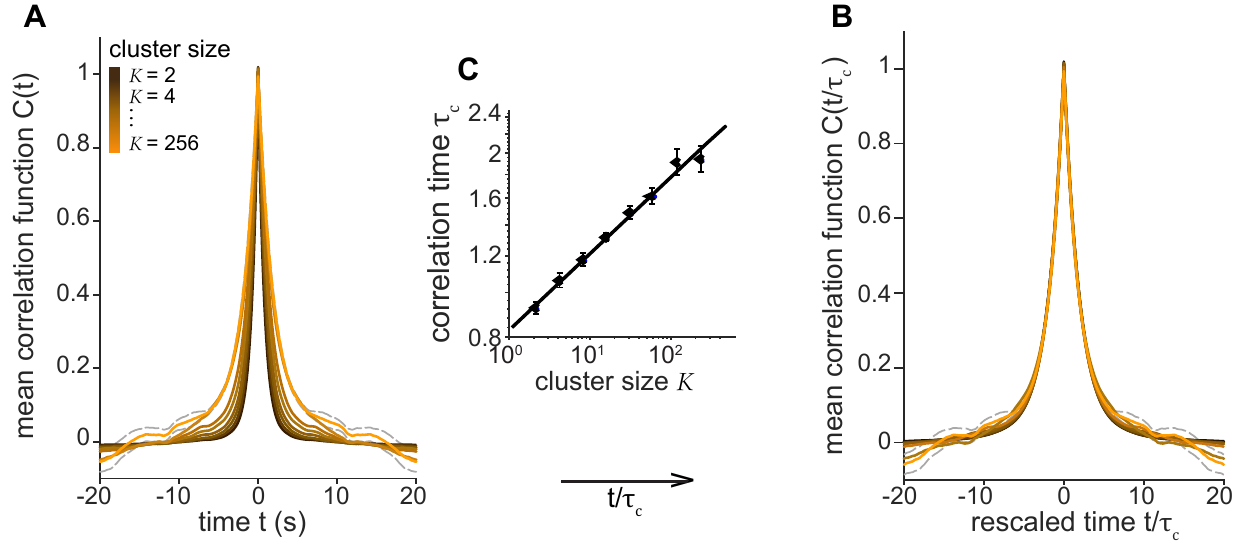}
	\caption{{\textbf{Dynamic Scaling.}}
		\textbf{(A)} Mean correlation functions for coarse--grained variables. Activity is summed in clusters of $K=2,\,, 4\, \cdots ,\, 256$ neurons (lightest orange corresponds to the largest cluster), with larger clusters exhibiting slower dynamics.  In dashed gray, $\pm$ one standard deviation across the population of $256$ neuron clusters. 
		\textbf{(B)} Collapse of curves by scaling of the time axis, $t\rightarrow t/\tau_c (K)$.  
		\textbf{(C)} Correlation time vs cluster size, fit to $\tau_c \propto K^{\tilde z}$, with $\tilde z = 0.16 \pm 0.02$
	\label{dynamic} 
	}

\end{figure*}

It is a common observation that fluctuations on larger spatial scales relax more slowly, and this idea is made precise by dynamic scaling relations.  Here we don't implement coarse--graining by spatial averaging, but we have the same intuition that the summed activity in larger groups of the neurons should exhibit slower dynamics.  In Figure \ref{dynamic}A we show the temporal correlation functions of the coarse--grained variables in groups of different size $K$.  The dynamics at different degrees of coarse--graining (Fig \ref{dynamic}A) indeed differ, but when we rescale by a single $K$--dependent correlation time the curves collapse (Fig \ref{dynamic}B).  The correlation time that effects this collapse scales as $\tau_c \propto K^{\tilde z}$, with $\tilde z = 0.16 \pm 0.02$, over two decades in cluster size $K$ (Fig \ref{dynamic}C).

\section{Results of coarse--graining via eigenmodes}

\begin{figure}[b]
\includegraphics[width=\linewidth]{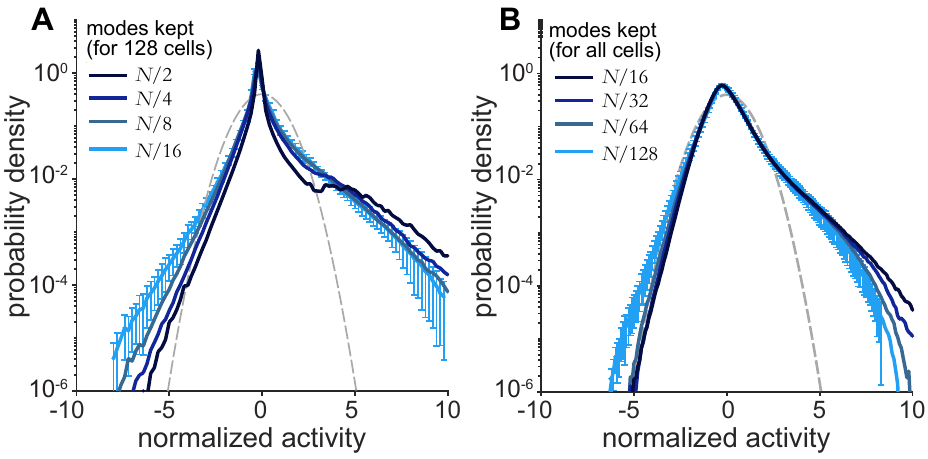}
\caption{\textbf{Scaling behavior for coarse graining in ``momentum'' space.} {\bf{(A)}} Probability distributions of activity, $P_{\hat K}(\phi)$. Clusters of $N=128$ neurons are analyzed after averaging over modes that make small contributions to the variance. We keep the top $N/2$, $N/4$, $N/8$, and $N/16$ modes (darkest to lightest blue, respectively); error bars are standard deviations across the different 128--cell clusters that we identify by coarse--graining in real space.
	{\bf{(B)}} Probability distributions of activity after coarse--graining in ``momentum'' space, $P_{\hat K}(\phi)$.  The full population of $N=1485$ cells, keeping the top $N/16$, $N/32$, $N/64$, and $N/128$ modes (darkest to lightest blue, respectively); error bars are standard deviations across random quarters of the experiment.
	\label{pca+rg}}
\end{figure}

In systems with translation invariance, we can implement coarse--graining either in real space or in Fourier space.  As explained above, projecting out the short wavelength (high momentum) Fourier modes is analogous to projecting out principal components that make small contributions to the total variance \cite{BraddeBialek2017}.  We are interested in what happens to the distribution of the resulting coarse--grained variables [Eq (\ref{k-spaceP})] as we change the scale of coarse--graining. Results are shown for clusters of $K=128$ neurons in Fig \ref{pca+rg}A.

We recall that, as we reduce the cutoff $\hat K$, the coarse--grained variables $\phi_{\hat K}({\rm i})$ [Eq (\ref{phiKi})] become weighted averages over more and more of the original variables.  If the correlations were weak, the central limit theorem would drive the distribution of these variables toward a Gaussian.  In fact  the distribution changes only very slowly as we move from $\hat K = K/2$ down to $\hat K = K/16$, perhaps even approaching a fixed non--Gaussian form.  To test this more fully we would like to have a wider range over which we can vary the cutoff $\hat K$, but this requires looking at larger populations of neurons.

With a limited number of samples, the eigenvalues of the covariance matrix for larger populations will become significantly distorted by sampling effects, but we don't actually need the eigenvalues themselves in order to implement coarse--graining. Rather what we need is that the structure and ordering of the eigenvectors is approximately correct, and much less is known about this problem than is known about the eigenvalues \cite{BunBouchaudPotters2017,MonassonVillamaina2015,Monasson2016}. So, recognizing that we need to be careful in interpreting the results, we simply try the same ``momentum shell'' coarse--graining procedure for the full population of $N=1485$ neurons.  The first thing we notice  in Fig \ref{pca+rg}B is that the probability distributions really are quite similar to what we saw in the better controlled case of $K=128$, although  a bit more rounded near $\phi =0$.  But now we can {\em start} at $\hat K = N/16$ and move down by another three factors of two, until we are keeping less than $1\%$ of the original degrees of freedom.  The distributions still are not Gaussian, and the difference in distributions when we change $\hat K$ by a factor of two is almost undetectable despite  small error bars.  Although we must be cautious because of the sampling problems, this is a strong hint that  this network of neurons is described by a non--Gaussian fixed point of the renormalization group transformation.

\begin{figure}[t]
	\includegraphics[width=\linewidth]{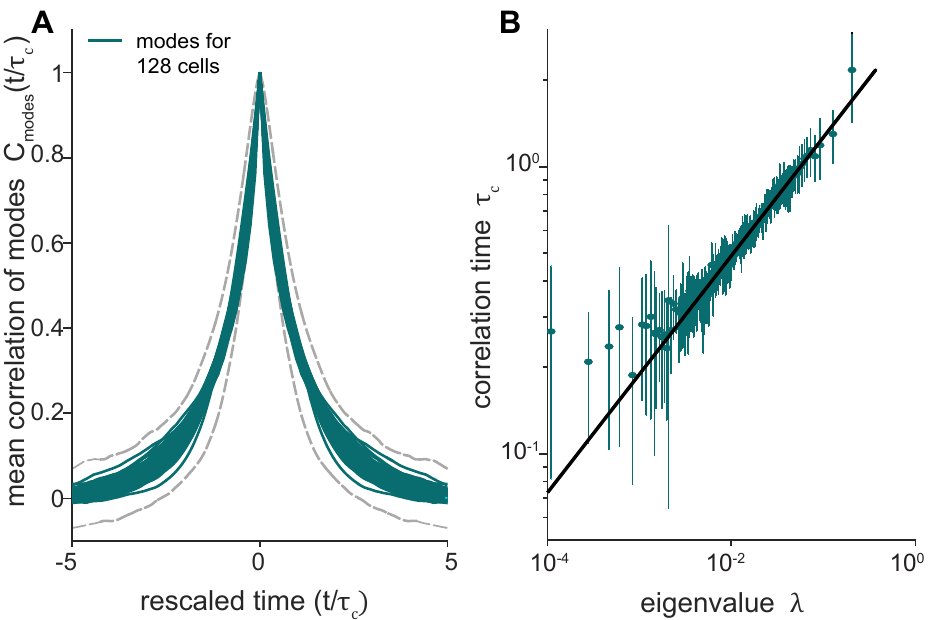}
	\caption{{\bf{Dynamic scaling for coarse graining in ``momentum'' space.} \bf{(A)}} Mean correlation functions for the modes in clusters of $K=128$ neurons, vs scaled time.  In dashed gray,  $\pm$ one standard deviation across the different clusters, for the mode the largest $\lambda$.  
	{\bf{(B)}} Correlation time vs eigenvalue, with fit to $\tau_c \propto \lambda^{z'}$, with $\tilde z = 0.37 \pm 0.04$.  
	\label{dynamic2}}
\end{figure}

As a complement to the ``real space'' observations of dynamic scaling, we consider the correlation functions of the different modes in groups of $K=128$ neurons, defined in Eq (\ref{modecorr}).   As in Fig \ref{dynamic}A, we choose this scale because then the number of independent samples is much larger than the dimensionality.  Correlation times for the different modes are spread over a full order of magnitude, but correlation functions collapse (Fig \ref{dynamic2}) and correlation times scale (Fig \ref{dynamic2}D) with $z' = 0.37 \pm 0.04$.    We note that the data are a bit scattered for the modes with small variance and short correlation time, but this is to be expected since these modes make very little contribution to the variance, and hence are most corrupted by noise, and the shortest correlation times come within a factor of two of the response time of the calcium indicator itself.   The ratio of dynamic exponents,  $z'/\tilde z = 0.52\pm 0.13$, is within experimental error of the exponent $\mu = 0.71\pm0.06$ describing the scaling of eigenvalues, as expected.

\section{Reproducibility}
\label{repro}

The inherent simplicity of scaling laws stands in contrast to the complexity and diversity exhibited by biological systems, and by the brain in particular.   In the successful applications of the renormalization group, an essential result is that scaling behaviors are invariant across a wide range of microscopic details, defining universality classes.  Before asking whether there are universality classes for the dynamics of real neural networks, we have to answer the more modest question of whether the scaling behaviors that we see are reproducible across multiple examples of the ``same'' network in different individuals.

For invertebrates, many neurons can be ``identified'' because they exhibit relatively stereotyped structure and functional behavior across all individuals in the same species \cite{bullock+horridge_65,WormAtlas,JefferisHummel2006,VosshallWongAxel2000}. While there are hints that even these identified  neurons can have properties that are specific to individuals \cite{schneidman+al_02,MurthyFieteLaurent2008,ColbertBargmann1995}, the whole picture is qualitatively different for vertebrates and especially for mammals such as the mice we study here (and us).  In mammalian cortex, neurons can be grouped into classes, and circuits are defined by statistical regularities in the connections among classes.  These statistical features of the network are thought to be reproducible across individuals, whereas detailed patterns of connectivity are not.   There are even significant individual differences in the number of cells that one finds in particular brain regions, including the hippocampus.

As a first test of whether the scaling behaviors that we see are invariant to the microscopic variability across individuals, we have analyzed the same experiment done independently in three mice.  In each case activity was recorded in the dorsal--posterior part of CA1 region in the hippocampus, with same experimental apparatus, as the mouse ran through the same virtual environment.  The number of neurons that we could identify in the experiment's field of view varied by $\sim 10\%$.  Qualitatively, all three data sets exhibit scaling in all of the features discussed above; quantitative results are summarized in Table \ref{exponents}.

\begin{table}[b]
	{\footnotesize
		\begin{tabular}{||c|c|c|c|c|c||}
			\hline
			$N$ &  $\tilde\beta$ & $\mu$ & $\tilde z$ & $\tilde z'$ &$\alpha$ \\
			\hline\hline
			$1485$ &  $0.88\pm 0.01$ & $0.71\pm 0.06$ & $0.16\pm 0.06$ & $0.37\pm 0.04$&$1.4\pm 0.06$\\
			\hline
			$1487$&  $0.89\pm 0.01$ &  $0.73\pm 0.01$ & $0.17\pm 0.03$ & $0.32\pm 0.17$& $1.56\pm 0.03$\\
			\hline
			$1325$&  $0.86\pm 0.02$ & $0.83\pm 0.07$ & $0.34\pm 0.12$ & $ 0.28\pm 0.13$&$1.73\pm 0.11$\\
			\hline\hline
	\end{tabular}}
	\caption{Exponents from three independent experiments. \label{exponents}}
\end{table}

The most precisely determined exponent is $\tilde\beta$, which describes the sub--extensive scaling of the effective free energy, determined by the probability of silence.  The errors in estimating $\tilde\beta$ are $\pm 0.014$ (root--mean--square across all three experiments), and the standard deviation across the independent measurements is $0.015$.  Thus, this exponent is reproducible within errors, and the errors are small.  In comparing the different exponents, it might be more fair to consider $\tilde\beta$ as a measurement of the deviation from naive extensive scaling, in which case what is significant is $1-\tilde\beta$, which is determined with $\sim 10\%$ accuracy.

\begin{figure}[b]
	\includegraphics[width = \linewidth]{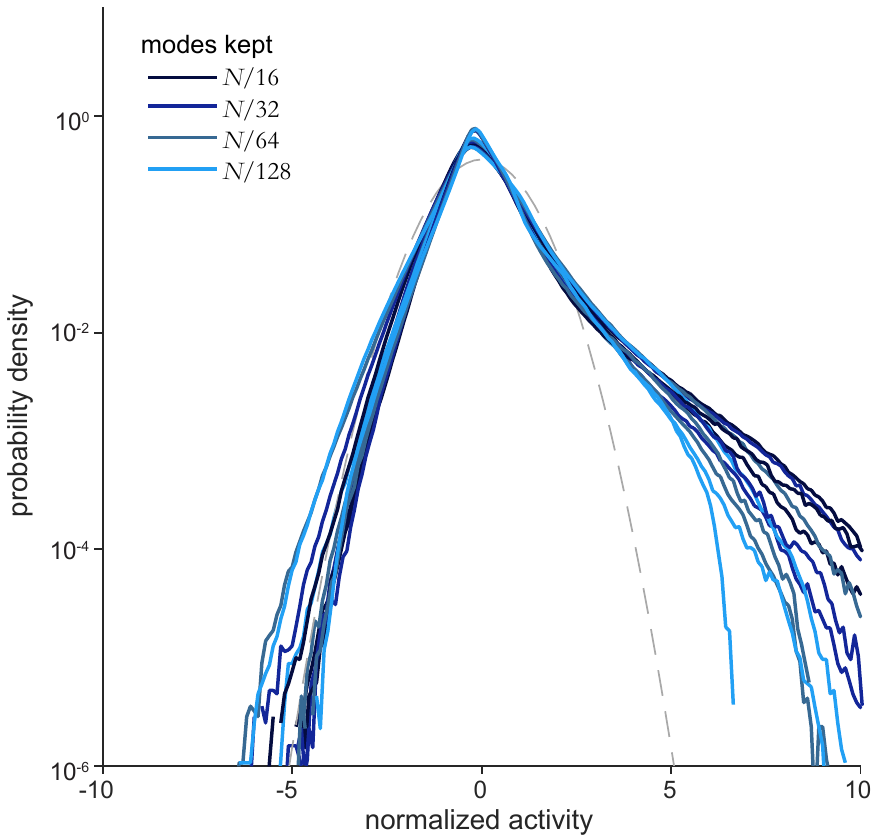}
	\caption{{\bf{Distributions of coarse--grained activity in ``momentum'' space for three independent experiments.}} The three overlaid probability distributions, $P_{\hat K}(\phi)$, were computed for the full population of cells, keeping the top $N/16$, $N/32$, $N/64$, and $N/128$ modes (darkest to lightest blue, respectively), as in Fig \ref{pca+rg}B. 
		\label{convergence}}
\end{figure}

The scaling of eigenvalues with rank also is determined with $\sim 10\%$ accuracy in single experiments, and reproducible with almost the same precision across experiments, as shown in Table \ref{exponents}.  Collecting all the data we have $\mu = 0.76\pm 0.05  \pm 0.06$, corresponding to the mean, the rms error of individual measurements, and the standard deviation across experiments.  Similar results are obtained for the dynamic scaling exponent from direct correlations,  $\tilde z = 0.22\pm 0.08 \pm 0.10$.  The measurement of dynamic scaling from eigenmodes has larger relative errors, but again we see agreement across experiments, $\tilde z' = 0.32 \pm 0.13 \pm 0.04$. Finally, the probability densities of the coarse--grained activity from different experiments converge to the same distribution, as shown in Fig \ref{convergence}.

The one exponent that does not fit this consistent pattern of reproducibility within error bars is $\tilde\alpha$, which describes the growth of the variance in activity with the size of the cluster. We find that variations across the three experiments are twice as large as expected from the errors in our measurements, $\tilde\alpha = 1.56 \pm 0.07 \pm 0.16$.  It is not clear to us whether this is significant or a fluctuation.  Certainly the preponderance of evidence is in favor of reproducibility, down to the $\sim 10\%$ precision with which individual exponents are determined.   

\section{Relations to place cell activity}

\begin{figure*}
	\includegraphics[width=\linewidth]{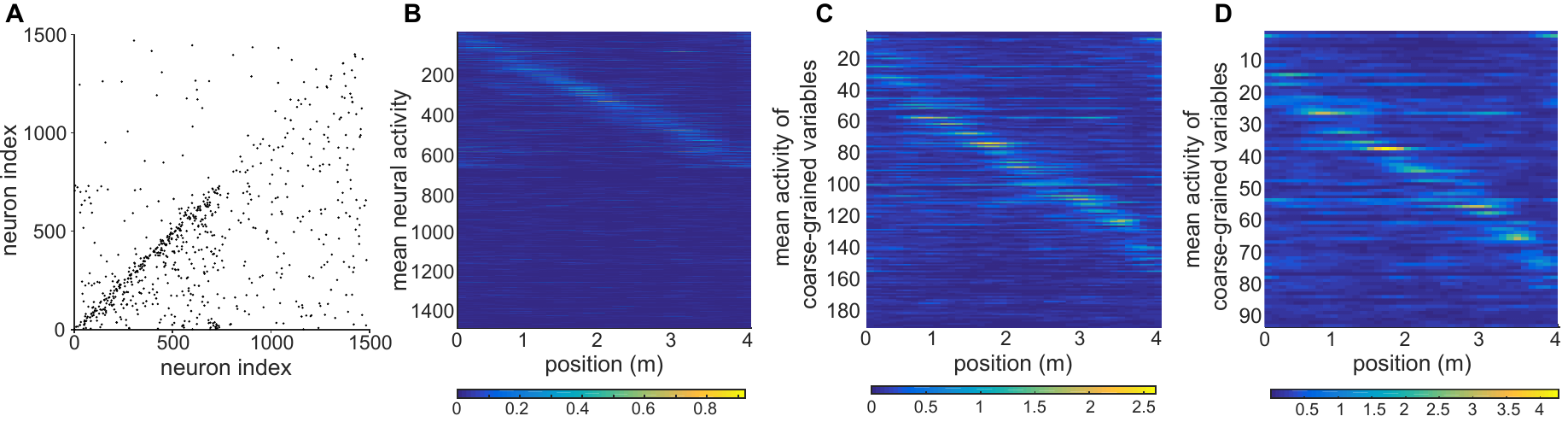}
	\caption{\textbf{Place cells behavior during coarse--graining.} 
		\textbf{(A)} Scatter plot of the pairing of neurons during the first iteration of the coarse--graining procedure. The first $\sim$ half of the neurons, 684 place cells out of 1485 total, tend to pair up with other place cells with a close field.
		\textbf{(B)} Averaged activity calibrated against positions, before coarse-graining. Place cells were sorted based on their activity's center of mass. Bottom $\sim$half are the non-spatially selective cells.
		\textbf{(C)} Averaged activity calibrated against position after 3 coarse-graining iterations, for $K=8$.
		\textbf{(D)} Averaged activity calibrated against position after 4 coarse-graining iterations, for $K=16$. 	
		\label{place_cells}
	}
\end{figure*}

The most salient qualitative fact about the hippocampus is the existence of place cells.  Roughly half of the neurons in the population that we observe fall into this category (Fig \ref{place_cells}), having near zero probability of begin active outside of a compact region along the virtual linear environment that the mice explore.  The other half of the cells could be place cells in a different environment, or might represent other variables \cite{LeutgebRagozzinoMizumori2000,GauthierTank2018,KrausRobinsonWhiteEtAl2013}.  What is the relationship between the map--like structure of place cell activity and the behavior that we see under coarse--graining?  There are, at least, two distinct questions here.  The first question is  whether what we see is consistent with place cell behavior.  As an example, if coarse--graining clustered together neurons that have very different place fields, then we might worry that our search for simplification is throwing away information of obvious relevance to the organism. The second question is whether what we see is a consequence of place cell behavior.  As an example, as the mouse runs the place cells are active in sequence, and one could imagine that this sequential pattern could generate something that at least approximates dynamic scaling.

The correlations that we use in defining our coarse--graining procedure encode information about the map--like structure of place cell responses, since cells with nearby place fields should have strong positive correlations in their activity. A closer look at the evolution of correlations (Fig \ref{evolution_corrs}) reveals that the first step of coarse--graining based on direct correlations tends to pair a place cell with another place cell that represents a nearby position (Fig \ref{place_cells}A).  As a result,  the map--like structure of average activity vs position persists for the coarse--grained variables, and this continues for several iterations, as demonstrated in Fig \ref{place_cells}B-D.   There is even a tendency for the a larger fraction of the coarse--grained variables to be spatially selective, as can be seen by comparing Fig \ref{place_cells}s B and D.  We conclude, at least qualitatively, that the place information known to be of relevance to the organism is preserved under coarse--graining.

The second question is whether place activity alone can account for the scaling behaviors we observe. To explore this issue we constructed a model of conditionally independent place cells \cite{MeshulamGauthierBrodyEtAl2017}.    For each neuron $\rm i$ we can estimate from the data the probability $p_{\rm i}(x)$ that the neuron will be active when the mouse is at position $x$ along the virtual track (Fig \ref{place_cells}B).  We then use the actual trajectories $x(t)$ form the data, and at each time $t$ we choose the states of neurons independently, $\sigma_{\rm i}(t) = 1$ with probability $p_{\rm i}(x(t))$.  We emphasize that this model of independent place cells reproduces, by construction, exactly the place fields exhibited by the real neurons, and inherits dynamics from the animal's movement through the (virtual) environment; while neurons are conditionally independent, they become correlated through the shared variable $x(t)$.

\begin{figure}
	\includegraphics{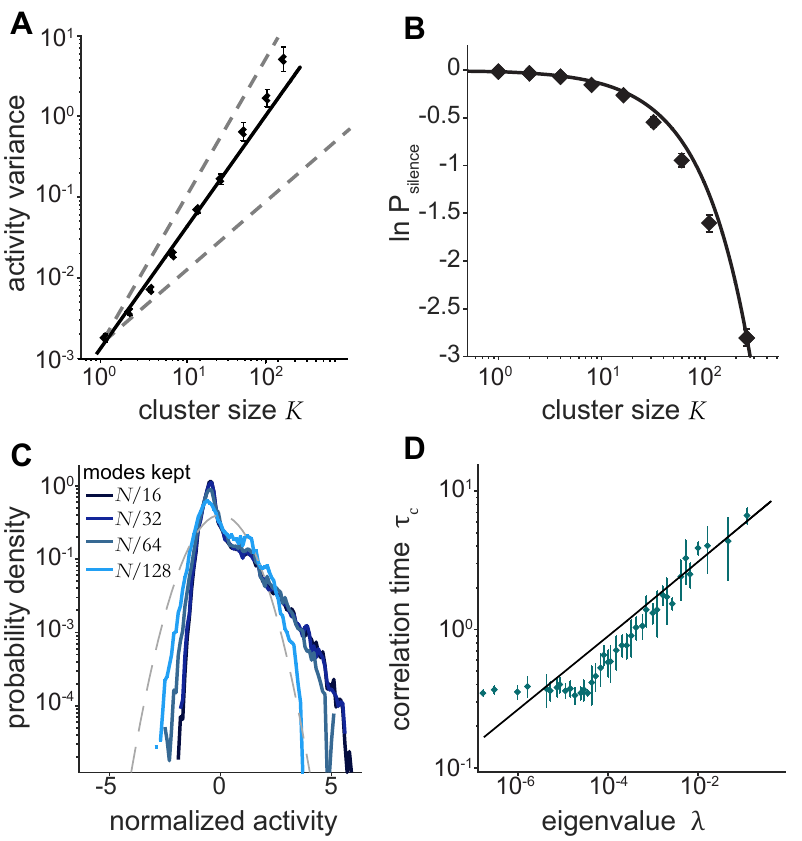} 
	\caption{\textbf{Failure of scaling in the independent place cell model.} \textbf{(A)} Variance of coarse--grained activity vs cluster size, from Eq (\ref{eqvar}).  Error bars estimated from random quarters of the data, shown at four times actual size. Black line indicates the best fit,  $M_2(K) \propto K^{\tilde\alpha}$, with $\tilde\alpha = 1.78 \pm 0.03$; dashed lines are $\tilde\alpha = 1$ and $\tilde\alpha =2$.  Should be compared with Fig \ref{var} for the real data.
		\textbf{(B)} Probability of silence in clusters of different sizes, cf Fig \ref{PsilencePact}A for the real data.
		\textbf{(C)} Probability distributions of activity after coarse--graining in ``momentum'' space, $P_{\hat K}(\phi)$.  The full population of $N=1485$ cells, keeping the top $N/16$, $N/32$, $N/64$, and $N/128$ modes (darkest to lightest blue, respectively). Should be compared with Figs \ref{pca+rg}B and convergence for the real data.
		\textbf{(D)} Correlation time vs eigenvalue  in ``momentum'' space. 
	} 
	\label{scaling_ind_place}
\end{figure}

We have simulated the independent place cell model, and analyzed the resulting data using the same coarse--graining procedures as used on the real data;  results are summarized in Fig \ref{scaling_ind_place}.  To begin, the variance of coarse--grained variables grows with the coarse--graining scale, but wanders  systematically around the best fit power--law relationship,  falling below the line at small $K$ and above the line at large $K$ (Fig \ref{scaling_ind_place}A).    The probability of silence falls with cluster size (Fig \ref{scaling_ind_place}B), but not quite with the scaling behavior of Eq (\ref{FvsK}).   These differences may seem small, but are much larger than the error bars, and should be compared with the nearly perfect scaling behaviors that we see in the real data (Figs \ref{var} and \ref{PsilencePact}A).  

We also analyze the independent place cell model via coarse--graining in ``momentum'' space.  In contrast with the real data (Fig \ref{pca+rg}), it is much less clear whether the probability distribution of the coarse--grained variables is approaching a fixed from as we project out a larger and larger fraction of the modes (Fig \ref{scaling_ind_place}C).  For the dynamics, we can achieve an approximate collapse of the correlation functions, but the correlation time does not scale with the corresponding eigenvalue (Fig \ref{scaling_ind_place}D). In smaller populations of neurons,  the independent place cell model fails to capture the quantitative behavior of pairwise correlations, either in calcium imaging data as analyzed here \cite{MeshulamGauthierBrodyEtAl2017} or in direct recordings of action potentials \cite{low+al_18}.

The results of Fig \ref{scaling_ind_place} provide convincing evidence that place cells are not an ``explanation'' of scaling behavior:  we can construct a population of neurons which has exactly the same pattern of place cells but does not scale.  This result perhaps should not be surprising.  In a population of place cells, there are two length scales, the approximate width of the place fields and the mean distance between place field centers.  The ratio of these lengths (in the one--dimensional environment studied here), gives us a characteristic number of neurons, which is $K_c \sim 18$ in our data.  we note that the plot of variance vs cluster sizes crosses the best fit power--law at $K\sim K_c$ (Fig \ref{scaling_ind_place}A), and  the probability of silence falls away from the best fit scaling behavior also at $K\sim K_c$ (Fig \ref{scaling_ind_place}B).  While these results are approximate, they highlight the fact that, in the presence of such obvious scales, the observation of rather precise power--law scaling in both static and dynamic quantities really is surprising.

\section{Discussion}

The renormalization group was developed as a framework for analyzing models, but it also provides inspiration for the analysis of experimental data.  The essential idea is that systematic coarse--graining should lead to simplification. Here we have used this idea to analyze experiments on the activity of 1000+ neurons  in mouse hippocampus.  To construct the analog of real space renormalization, we have taken as ``neighbors'' the neurons whose activity is maximally correlated, and to construct the analog of momentum space renormalization we have used the mapping between momentum shells and the eigenmodes of the covariance matrix.  With both approaches, we observe quite precise scaling behaviors of static and dynamic quantities across two decades.   Moreover, the results are highly reproducible across multiple experiments, with scaling exponents agreeing within error bars, at the $10\%$ level or better.   These results suggest that there is an underlying, non--trivial fixed point of the RG that would provide a simpler description of collective network activity.

Because our coarse--graining procedure follows the strongest correlations along a greedy path, one might worry that it is especially sensitive to spurious correlations which arise from the finite size of our data set.  Thus it is important that all the non--trivial scaling behaviors disappear when we shuffle the data (Appendix \ref{shuffle}).   In the same spirit, we have checked that generic recurrent networks, which can have interesting dynamics \cite{vogels+al_05}, do not exhibit scaling (Appendix \ref{rnn}).  Perhaps more deeply, we can construct populations of neurons that reproduce exactly the pattern of place cells seen in the hippocampal data, but these model networks also don't exhibit scaling (Fig \ref{scaling_ind_place}).  We conclude that the scaling behaviors we see are not artifacts of limited data, that they are not the behaviors of typical neural networks, and that they are not simple consequences of the known hippocampal representation of space.  

The search for scaling behaviors in biological systems has a long history.   Allometric scaling, which concerns the variation of functional parameters with body size, and the size relations of different body parts to one another, goes back at least to the 1890s and was codified by Julian Huxley in the 1930s \cite{huxley_32}.  Following the dramatic success of the RG in understanding critical phenomena, there has been a much wider search for scaling behaviors in complex, non--equilibrium systems, including biological systems.   Controversy surrounds much of this work, not least because of the difficulties of convincingly identifying power--law behaviors in limited data sets \cite{ClausetShaliziNewman2009}.  

The renormalization group predicts that, if we have non--trivial scaling behaviors, then these are asymptotically exact, reproducible, and even universal.    But there is more to scaling than power laws.  We should see that probability distributions of  coarse--grained variables approach fixed non--Gaussian forms, and we should see collapse of correlation functions.  Although our data are limited, not least because we are looking only at $1000+$ neurons, we see these signatures of scaling.  We are particularly struck by the precision and reproducibility of the scaling behaviors.  Because neural activity is sparse, one of the things we can measure best is the probability that a population of cells is silent.  Within a statistical mechanics description [Eqs (\ref{boltz1}, \ref{boltz2})], this probability of silence translates directly into an effective free energy, and we find that this free energy scales very precisely across more than two decades (Figs \ref{PsilencePact}A and \ref{scaling_cont}A).  The exponent that describes this scaling is the one that we estimate most accurately, and it is reproducible within these very small error bars, $\tilde\beta = 0.87\pm 0.014\pm 0.015$.

Our coarse--graining procedure is based on equal--time correlations, so the observation of dynamic scaling seems especially significant.  In particular, the analysis in the space of eigenmodes shows that correlation functions collapse and correlation times scale with the eigenvalue, with no sign of saturation (Fig \ref{dynamic2}B).  If this pattern holds in larger populations of cells, it means that neural networks have access to a continuum of time scales, with modes that fluctuate on collective time scales much longer than the characteristic times of individual neurons, limited only by the size of the network.  This emergence of long time scales may be the most important functional consequence of scaling.

Our results on scaling in a neuronal population are  closely connected to observations on scaling in collective animal behavior.  In flocks of birds, swarms of insects, and populations of bacteria, a variety of static and dynamic scaling behaviors have been demonstrated \cite{CavagnaContiCreatoEtAl2017,CavagnaCimarelliGiardinaEtAl2010,ChenDong2012}.  In all these cases, as with our results here, the observation of scaling is just a first step, and it remains a challenge to identify the theory which describes the underlying RG fixed point.

We emphasize that the approaches we have taken here are in many ways only a first step.  As emphasized at the outset, the coarse--graining step of the RG is not unique, and there are many possibilities.  In the familiar examples from statistical mechanics, symmetries often constrain the form of the coarse--graining transformation, but for neurons we have more flexibility.  We can imagine compressing the state of $K$ neurons into a coarse--grained variable not simply by averaging or majority rule, but by some more subtle transformation that would preserve information about sensory inputs, motor outputs, or even the future state of the network \cite{palmer+al_15}.  Yet another possibility\footnote{We thank MO Magnasco for this observation.} is to assign each neuron coordinates $\mathbf{r}_{\rm i}$ in some $D$ dimensional space such that the correlations $c_{\rm ij}$ [Eq (\ref{cij_def})] are a (nearly) monotonic function of the distance $d_{\rm ij} = | \mathbf{r}_{\rm i} - \mathbf{r}_{\rm j}|$; this is multidimensional scaling \cite{kruskal_64}.   Our coarse--graining procedure then starts as a local averaging in this abstract space, and it would be interesting to take this embedding more seriously as way of recovering locality of the RG transformation.

It is not guaranteed that coarse--graining, in the spirit of the RG, will lead to non--trivial results.  As we see in Appendix \ref{rnn}, interesting models of neural network dynamics do not exhibit clear scaling behavior, and coarse--grained variables seem to be described by Gaussian distributions.  In equilibrium statistical mechanics, coarse--graining leads to non--trivial fixed points only at critical values of the underlying parameters.  In this sense, our results here are connected to previous ideas about criticality in neural networks, and in biological systems more generally \cite{mora+bialek_11}.  One version of this idea makes an analogy between  ``avalanches'' of sequential activity in neurons \cite{beggs+plenz_03,friedman+al_12} and  the early sandpile models for self--organized criticality \cite{bak+al_87}.  A very different version focuses on the distribution over microscopic states at a single instant of time  \cite{tkacik+al_15}, and is more closely connected to criticality in equilibrium statistical mechanics.     In our modern view, invariance of probability distributions under iterated coarse--graining---a fixed point of the renormalization group---may be the most fundamental test for criticality, encompassing both static and dynamic scaling,  independent of analogies to thermodynamics.

A  fundamental result of the renormalization group is the existence of irrelevant operators, which means that successive steps of coarse--graining lead to  simpler and more universal models.  Although the RG transformation begins by reducing the number of degrees of freedom in the system, as we have emphasized simplification does not result from dimensionality reduction but rather from the flow through the space of models.  The fact that a phenomenological approach to coarse--graining  of neural data gives results which are familiar from successful applications of the RG in statistical physics encourages us to think that simpler and more universal theories of neural network dynamics are possible.

\begin{acknowledgments}
We are grateful to S Bradde, A Cavagna, R Engelken, I Giardina, MO Magnasco, SE Palmer, and DJ Schwab for many helpful discussions.  This work was supported in part by the National Science Foundation through the Center for the Physics of Biological Function (PHY--1734030), the Center for the Science of Information (CCF--0939370), and grant PHY--1607612; by the Simons Collaboration on the Global Brain; and by the Howard Hughes Medical Institute.
\end{acknowledgments}

\vfill\newpage
\appendix

\section{Controls}
In conventional statistical physics problems, the appearance of power--law scaling and the approach of probability distributions to a fixed form provides unambiguous evidence that the system we are studying is described by a fixed point of the RG.  But the classes of models that we are allowed to consider in these problems are highly constrained, by the locality of interactions, symmetries, and more.  In the more complex biological systems that we are interested in here, one might worry that seemingly compelling signature could arise for very different reasons.  

\subsection{Absence of scaling in shuffled data}
\label{shuffle} 

\begin{figure}
	\includegraphics[width = \linewidth]{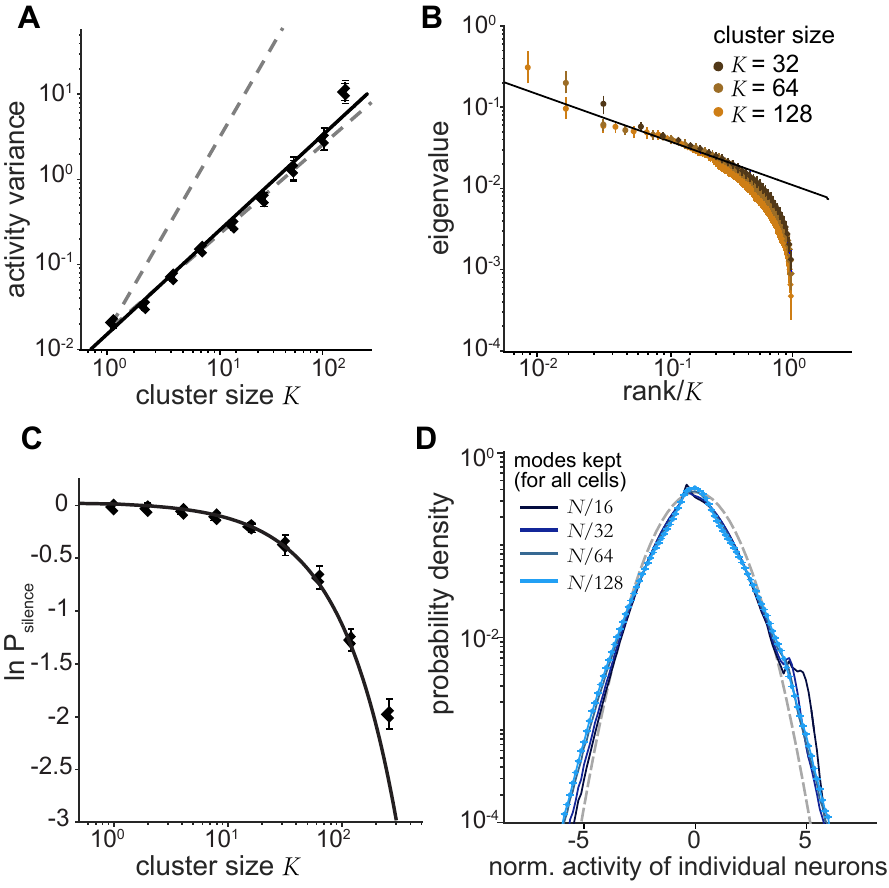}
	\caption{\textbf{Absence of scaling in shuffled data.} 
		\textbf{(A)} Variance of coarse--grained activity vs cluster size, from Eq (\ref{eqvar}), for shuffled data; compare with Fig \ref{var} for real data.  Error bars estimated from random quarters of the data, shown at four times actual size.  Dashed lines show $\tilde\alpha = 1$ and $\tilde\alpha =2$, for comparison. Black line indicates the power--law fit, which is very close to the $\tilde\alpha = 1$, as expected from non--correlated samples.
		\textbf{(B)} Eigenvalues of the covariance matrix [Eq (\ref{Cij})] for clusters of size $K=32$ (dark brown), $64$ (mid brown), and $128$ (light brown); compare with Fig \ref{spectra} for real data.  Error bars are standard deviations across both random quarters of the experiment and the different clusters in our data set.  The fit to a  power--law  (black line) is poor, and we do not see collapse of data from clusters of different sizes.
		\textbf{(C)} Probability of silence in clusters of different sizes; compare with Fig \ref{PsilencePact}A for real data.  Error bars are standard deviation across random quarters of the data, and the curve is Eq (\ref{FvsK}) with $\tilde\beta = 1$.
				\textbf{(D)} Probability distributions of activity after coarse--graining in ``momentum'' space, $P_{\hat K}(\phi)$; compare with Fig \ref{pca+rg} for real data.  The full population of $N=1485$ cells, keeping the top $N/16$, $N/32$, $N/64$, and $N/128$ modes (darkest to lightest blue, respectively); error bars are standard deviations across random quarters of the experiment, and the dashed line is a Gaussian.
	} 
	\label{scaling_shuffle}
\end{figure}

Our effort to coarse--grain the states of a neural population starts with the pairwise correlations that we observe among the activity of neurons.    While the duration of our experiments is long enough that individual correlation coefficients can be determined with reasonable accuracy, one might worry that we do not have enough independent samples to make reliable statements about the global structure of the correlations among all $N=1485$ variables.     More specifically, we want to exclude the possibility that what we find is being driven by the spurious correlations that arise from the finite size of our data set.

To address concerns about spurious correlations, we construct a surrogate data set which breaks the real (equal time) correlations between neurons but has the same number of independent samples.  We note that a complete shuffling of the data---independently permuting the time indices for each neuron---is equivalent to the assumption that we have as many independent samples as we have time points, which surely is wrong.  Instead we rigidly shift the time series for each neuron by an independent random offset, so that temporal correlations are preserved but equal time correlations between neurons should be broken; this is also how we generate the randomized data for Fig \ref{corrs}A.  We then analyze these surrogate data exactly as with the real data; results are summarized in Fig \ref{scaling_shuffle}.

We expect that, with no correlations, the variance of the coarse--grained variables should grow linearly with $K$, and this is what we see (Fig \ref{scaling_shuffle}A); the $K=256$ point is a bit off, presumably the effects of spurious correlations are most serious here.   The spectra of correlations within a cluster show no evidence of scaling, having neither a clear power--law dependence on rank nor a collapse across different cluster sizes (Fig \ref{scaling_shuffle}B).  The probability of silence in the cluster falls linearly with cluster size (Fig \ref{scaling_shuffle}C), corresponding to $\tilde\beta = 1$, again with a small systematic deviation at the largest cluster sizes.  Finally, when we coarse--grain via eigenmodes, the distribution of coarse-grained variables converges quickly on a Gaussian (Fig \ref{scaling_shuffle}D), as expected from the center limit theorem.

To summarize, {\em none} of the signatures of scaling are present in our surrogate data set, and hence none of what we see can be ascribed to artifacts from finite sample size.

\subsection{Absence of scaling in a generic network}
\label{rnn}

Our intuition from equilibrium statistical mechanics is that scaling and the appearance of a non--trivial fixed point require some special tuning of parameters.  But perhaps in complex, non--equilibrium systems such as neural networks, we can see these effects more generically.  To test this possibility we consider a model neural network, following Ref \cite{vogels+al_05}.

\begin{figure}
	\includegraphics[width = \linewidth]{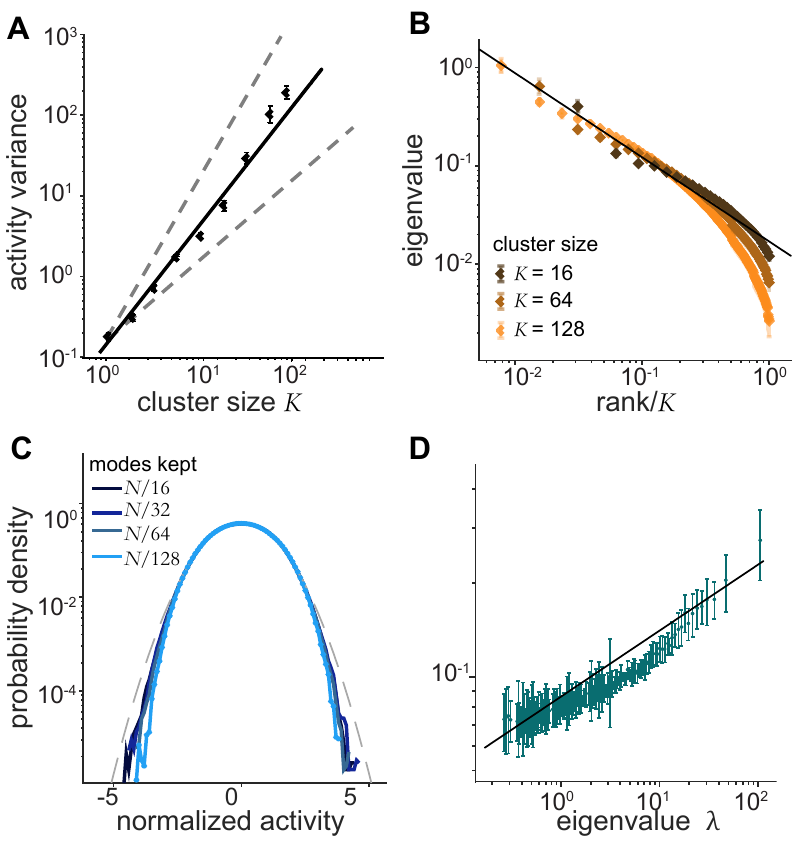}
	\caption{\textbf{Absence of scaling in a generic network.} 
		\textbf{(A)} Variance of coarse--grained activity vs cluster size, from Eq (\ref{eqvar}), for the simulation of a generic network; compare with Fig \ref{var} for real data.  Error bars estimated from random quarters of the data, shown at four times actual size.  Black line indicates the best power-law fit with $\tilde\alpha = 1.54 \pm 0.48$, but the data points fail to follow the line; dashed lines are $\tilde\alpha = 1$ and $\tilde\alpha =2$. 
		\textbf{(B)} Eigenvalues of the covariance matrix [Eq (\ref{Cij})] for clusters of size $K=32$ (dark brown), $64$ (light brown), and $128$ (orange); compare with Fig \ref{spectra} for real data.   Error bars are standard deviations across both random quarters of the data from the simulation and the different clusters.  
		\textbf{(C)}  Probability distributions of activity after coarse--graining in ``momentum'' space, $P_{\hat K}(\phi)$; compare with Fig \ref{pca+rg} for real data.   We analyze the full population of $N=1485$ cells, keeping the top $N/16$, $N/32$, $N/64$, and $N/128$ modes (darkest to lightest blue, respectively). Distributions converge to a Gaussian(dashed gray).
		\textbf{(D)} Dynamic scaling in ``momentum'' space; compare with Fig \ref{dynamic2}B for real data.  Correlation times deviate strongly  from the best fit power--law (black).
	} 
	\label{scaling_rnn}
\end{figure}

Each of the $N$ neurons in the network is described by a  real number $x_{\rm i}$, which in the absence of interactions would relax to zero on a time scale $\tau$.  Each neuron $\rm i$ can be driven by a nonlinear function of the variable describing neuron $\rm j$, and the strengths of these interactions are drawn randomly, to characterize the generic behavior.  Concretely,
\begin{equation}
\tau\frac{d\mathbf{x}}{dt} = -\mathbf{x}+\sum_{\rm j} J_{\rm ij} \tanh(x_{\rm j}),
\label{rnn_eq}
\end{equation}
where the $J_{\rm ij}$ are independent Gaussian random variables with $\langle J^2\rangle = g/N$.  We take $g=3$, which guarantees that the quiescent state of the network is unstable, and $\tau = 10\,{\rm ms}$ to set a time scale.  We choose $N=1485$, as in our data, and initialize the network in a random state.  The simulation generates a long time series, and we analyze these data as we did the continuous fluorescence signals \cite{meshulam+al_18}; results are summarized in Fig \ref{scaling_rnn}.

We see in Fig \ref{scaling_rnn}A that the variance of the coarse--gained variables does not scale with cluster size.  Instead we see a systematic variation around the best fit power--law, much as with the independent place cell model in Fig \ref{scaling_ind_place}.  The spectra of correlations inside the clusters (Fig \ref{scaling_rnn}B) does not show much evidence fo power--law behavior, and perhaps more significantly does not exhibit collapse across clusters of different sizes.  When we pass to the eigenmodes, coarse--graining leads quickly to a Gaussian distribution (Fig \ref{scaling_rnn}C).
Correlation functions for the different eigenmodes have significantly different shapes, and if we define a correlation time for each mode we do not see scaling with the eigenvalue (Fig \ref{scaling_rnn}D), as we do with the real data (Fig \ref{dynamic2}B).

To summarize, {\em none} of the scaling signatures that we see in the real data are present in the dynamics of a generic neural network.

\bibliography{scaling}

\end{document}